\documentclass[a4paper,12pt]{JHEP} % 10pt is ignored!

\newcommand\fverb{\setbox\pippobox=\hbox\bgroup\verb}
\newcommand\fverbdo{\egroup\medskip\noindent%
			\fbox{\unhbox\pippobox}\ }
\newcommand\fverbit{\egroup\item[\fbox{\unhbox\pippobox}]}
\newbox\pippobox

\newcommand{\be}{\begin{equation}}
\newcommand{\ee}{\end{equation}}
\newcommand{\bea}{\begin{eqnarray}}
\newcommand{\eea}{\end{eqnarray}}
\newcommand{\dalpha}{\dot{\alpha}}
\newcommand{\dbeta}{\dot{\beta}}
\newcommand{\dA}{\dot{A}}
\newcommand{\dB}{\dot{B}}

\font\mybb=msbm10 at 12pt
\font\mybbb=msbm10 at 8pt
\def\bb#1{\hbox{\mybb#1}}
\def\bbb#1{\hbox{\mybbb#1}}

\def\zet{{\bb{Z}}}
\def\zzet{{\bbb{Z}}}
\def\real{{\bb{R}}}
\def\rreal{{\bbb{R}}}

\def\R4{\real^4}
\title{ D-instanton probes of ${\cal N}=2$ non-conformal geometries} 
\author{Francesco Fucito
\\
INFN sez. di Roma 2,\\
Via della Ricerca Scientifica, 00133 Roma, Italy\\
E-mail: \email{fucito@roma2.infn.it}}
\author{Jose F. Morales
\\
Spinoza Institute, Utrecht, The Netherlands\\
E-mail: \email{morales@phys.uu.nl}}
\author{Alessandro Tanzini
\\
Laboratoire de Physique Th\'eorique et Hautes Energies\\
 Universit\'e Paris~VI, 4 Place Jussieu 75252 Paris Cedex 05, France\\
and INFN, Roma, Italy.\\
E-mail: \email{tanzini@lpthe.jussieu.fr}}
\preprint{\hepth{0106061}}
\abstract{
D-instanton calculus has proved to be able to probe the 
AdS near horizon
geometry for $N$ D-branes systems which, when decoupled from gravity,
yield four dimensional superconformal gauge
theories with various matter content.
In this work we extend
previous analysis to encompass fractional brane models which
give rise to non conformal ${\cal N}=2$
Super Yang-Mills theories.
Via D-instanton calculus we study the geometry of such models 
for finite $N$ and recover
the $\beta$ function 
of the gauge coupling constants which is expected 
in non conformal gauge theories. 
We also give a topological matrix theory formulation
for the D-instanton action of these theories.
Finally, we revisit the related system 
where the D3-branes wrap a ${\real}^4/{\zet}_p$
orbifold singularity and the D(-1) branes are associated to instanton 
solutions of four-dimensional gauge theories in the blown down ALE space.}

\keywords{Instantons, RG-flows, Non-conformal gauge theories, ALE spaces}
\begin{document}
\setcounter{footnote}{0}

\section{Introduction}
The present work arises from two distinct sets of motivations which
point both in the direction of D-branes.

The first motivation is given by the difficulties that one finds
in extending the Maldacena's conjecture \cite{Maldacena:1998re} to
${\cal N}=2$ models which do not enjoy the property of
conformal invariance \cite{fgpw}-\cite{bdflmp}. The classical
supergravity solutions dual to such non conformal gauge theories
typically possess infrared singularities which have to be
resolved to yield sensible models. A proposal for a resolution
of such singularities is known as the {\it enhan\c{c}on} mechanism: if a
massive probe moves in the backgrounds of the fractional 
D3-brane, it becomes tensionless before reaching the singularity. 
The supergravity description breaks down and one is forced to consider
stringy effects which presumably would smooth out the singularity.
Despite some recent progresses \cite{prz} in the understanding of 
the physics of the enhan\c{c}on, a complete satisfactory picture 
is still missing.

In the conformal versions
of the Maldacena's conjectures D-instanton calculus 
has been proved to
efficiently probe the geometry of the associated supergravity solution
\cite{dhkmv}. 
Indeed instanton
dominated correlators for the four
dimensional Yang-Mills theory (SYM$_4$) with color gauge group
$SU(N)$ and sixteen supersymmetric charges have been put in 
correspondence in the $N\to\infty$ limit with D-instanton corrections 
to higher derivative terms in type IIB supergravity on 
$AdS_5\times S^5$ \cite{dhkmv,bgkr}.  
The result relies on the fact, proved in \cite{dhkmv},
that the multi-instanton moduli space of ${\cal N}=4$ $SU(N)$
SYM$_4$ factors out, in the large $N$ limit, in an 
$AdS_5\times S^5$ term describing the overall
center of mass degrees of freedom and an effective
$SU(k)$ matrix model describing degeneracies of the 
bound state of $k$ D-instantons in ten dimensions.
Generalizations to other SYM$_4$ with different gauge group or
matter content have been addressed in \cite{gns,hkm,hk}.
In \cite{gns} the authors studied a model arising
from a system of $N$ D3-branes living at an orientifold 7-plane
together with eight D7-branes \cite{fsafm} 
that, in the low energy
limit and when gravity is decoupled, gives rise to superconformal
${\cal N}=2$ SYM$_4$ with
gauge group $Sp(N)$.
Depending on whether one tests the geometry with ``regular''
or ``fractional''
\footnote{By a ``fractional'' brane in the unoriented context we
mean an unpaired D-brane constrained to live on
the orientifold plane. A ``regular'' brane can move off the fixed point 
of the orientifold if it does not break the modding discrete symmetry.}  
D(-1)-instantons, the expected large $N$ geometry $AdS_5\times S^5/\zet_2$ 
or $AdS_5\times S^3$ were recovered respectively.
The case of IIB
string theory compactified to six dimensions on K3 with a
vanishing two cycle was coped with in \cite{hkm}. The low energy
theory is ${\cal N}=2$ SYM$_4$ with gauge group $SU(N)$ and
$N_F=2N$ fundamental hypermultiplets. The geometry seen by the instanton
probe was proved to be $AdS_5\times S^1$ in the large $N$ limit.

A more general class of models is defined by locating a stack
of D3-branes at a $\real^6/\Gamma$ singularity, with $\Gamma$ a
discrete subgroup of $SU(4)$. From the gauge theory point of view 
such moddings act on the ${\cal R}$-symmetry group
and with a suitable prescription they
reduce the number of supersymmetric charges without spoiling
the conformal invariance of the theory \cite{kasi}.
The resulting low energy
theory is ${\cal N}=0,1,2$ SYM$_4$ (depending on whether $\Gamma$ 
is embedded in a $SU(4)$, $SU(3)$ or $SU(2)$ subgroup of 
the  $SU(4)_{\cal R}$ ${\cal R}$-symmetry group respectively), 
with product gauge groups and
bifundamental matter. From the point of view of
instanton calculus, such theories, for $\Gamma=\zet_p$, were studied
in \cite{hk} where the near horizon geometry, $AdS_5\times
S^5/\zet_p$, was recovered in the large $N$ limit.

One of the aims of this paper will be the study, via
a {\it finite} $N$ D-instanton
calculus, of a non conformal variant of the models we just described,
involving both regular and fractional branes.
We focus ourselves on the ${\cal N}=2$ cases but most of our 
analysis extends to more general situations.
%How to extend
%instanton calculus to deal with these models will be the subject
%of section 3.
The deviations from conformal invariance seen from the
probe point of view will be shown to be
entirely determined by the one-loop $\beta$-function coefficients
of the underlying gauge theory.
The analysis of non trivial RG-flows in the context of 
non conformal ${\cal N}=2$ AdS/QFT correspondences 
of the kind we study was pioneered in
\cite{kn}.    
The subsequent results \cite{aa,bdflmp,bimo} strongly 
motivate the present work. 
In \cite{aa} the correspondence between the
one-loop beta function coefficient and the tadpole of the
dilaton in the bulk supergravity theory was tested in 
an ${\cal N}=0$ SYM$_4$ gauge theory constructed out of D-branes and
O-planes.
The result was extended in \cite{bimo} to various ${\cal N}=0,1,2$
non-conformal geometries. In this reference the one-loop
beta function coefficient is read from the   
annulus/cylinder string computation which is shown to be 
protected from contributions coming from 
massive string modes, admitting therefore 
equivalent description in terms of either supergravity  
or SYM$_4$ degrees of freedom. 
In \cite{bdflmp} the authors constructed
the explicit supergravity solution of a fractional
D3-brane outside of the enhan\c{c}on and showed how the expected   
running of the gauge coupling constant is recovered from it. 
In this work we study the ${\cal N}=2$ 
RG-flow from the D-instanton probe perspective.  

The second motivation for this paper was induced by the desire to
find a physicist's handle to some beautiful mathematical properties 
enjoyed by 
the moduli space of gauge connections realized {\it \`a la} ADHM
\cite{adhm}. Our present poor understanding of such properties
accounts, in our opinion, for the difficulties we have in
computing non perturbative effects in  SYM$_4$ theories.
This task turns out to
be one of the hardest challenges of present days theoretical
physics. In spite of some spectacular progresses \cite{sw}, the
only tool to study non perturbative properties in the framework
of field theory is via instanton calculus. The single hurdle that
makes such computations so difficult is given by the presence of
constraints in the ADHM construction of the moduli space of gauge
connections. Such constraints can be explicitly solved only
in a very limited number of instances but even in those cases the
amount of algebra involved in the computations tends to obscure
its physical meaning. To overcome this difficulty it has been
proposed \cite{Dorey:1998xb} to implement such constraints in the
functional integration via the introduction of a suitable number
of Dirac deltas. Given that this prescription reproduces the
correct measure (and it does \cite{Bruzzo:2000di}), we can now
write the instanton measure for arbitrary winding number. 
This strategy has proved very successful
especially in giving non perturbative checks \cite{dhkmv} of the
Maldacena's conjecture \cite{Maldacena:1998re}. In this last
instance, the absence of a vacuum expectation value for the
scalars of the theory and the large $N$ limit
conspire to give the desired result. Such combination of favorable
events is not always present. 
For example, giving a non trivial v.e.v. to the scalars corresponds 
in the brane picture to separate the branes, and      
this gives rise to less symmetric configurations rendering the
computations much more involved. 
This is why some of the authors have
proposed to exploit the topological properties of SYM$_4$
\cite{topo}, an approach that we extend here also to the
various ${\cal N}=2$ models arising from the above mentioned 
brane configurations; an extension to ${\cal N} =4$ has 
been carried out in \cite{dhk}.
% see more details later)
The localization
properties which show up in the topological formulation
give indeed useful simplifications.  
One of the most striking outcome in the first checks
\cite{Fucito:1997ua} of the non perturbative behaviour of ${\cal N}=2$
SYM$_4$ computed by Seiberg and Witten was in fact its agreement with
semiclassical instanton calculus. 
This in turn implies that the saddle point
approximation for ${\cal N}=2$ SYM$_4$ is exact: a typical feature of
topological theories. 
Along this line it is
possible to show that the relevant correlators can be written as a
total derivative in the moduli space of gauge connections: only
zero size instantons contribute to the computation. The measure is
then recovered as a change of coordinates from fermionic to
bosonic variables in the supersymmetric extension of the
hyperk\"ahler quotient construction \cite{Bruzzo:2000di}. But
where does the property of being hyperk\"ahler come from? Formally
it can be obtained considering the infinite dimensional
hyperk\"ahler quotient of the space of general connections by the
triholomorphic action of the group of gauge transformations ${\cal
G}$. 
The zero level set is given by the space of self-dual solutions which,
quotiented by ${\cal G}$, yields the moduli space. A much more
palatable derivation can be given using D-branes. The ADHM
construction for SYM$_4$ with sixteen supercharges, gauge group
$SU(N)$ and arbitrary instanton number $k$ can indeed also be
obtained from the low-energy effective action of 
a system formed by a stack of $k$ D$p$-branes
and $N$ D$(p+4)$ branes reduced to zero
space-time dimensions when gravity is switched off
\cite{wttn,Douglas:1998uz,dhkmv}. 
In this set up the familiar hyperk\"ahler properties of the 
ADHM instanton moduli space descends naturally from similar
structures in the underlying D(-1)--D3 supersymmetric gauge
theory (see \cite{Craps:1997gp} for a review from the field
theory perspective).
%The physicist's explanation we
%were looking for now comes from well known properties of
%supersymmetric theories, see \cite{Craps:1997gp} for a review.
Such intimate relationship between the ADHM construction and
D-branes lies also at the origin of the results which we commented
before relating moduli spaces of gauge connections and near
horizon $AdS$ geometry.

This is the plan and summary of the paper: in section 2 we collect
the main background material needed in the rest of the
paper. Section 3 is devoted to the study of the geometry associated
to various non conformal ${\cal N}=2$ gauge theories living on 
D3-brane systems located at $\real^4/\zet_p$ singularities. As usual, the tested 
geometry depends on whether we use fractional or regular
instanton probes. In the former case we found a geometry which
is a sort of non conformal deformation of $AdS_5\times S^1$.   
The position in the $AdS_5$ space is identified with the
energy scale of the boundary gauge theory and
the $\beta$-function coefficients agree with the field
theory expectations. 
In the case where we use regular instantons as a probe, the geometry
turns out to be $AdS_5\times S^5/\zet_p$ independently on whether
the theory is conformal or not. This is in agreement with
our expectations from both supergravity and field theory 
point of view, since, as we will see, regular instantons 
test only the overall sum of the coupling constants in the product 
gauge group, and it is precisely this combination that does not run.  
We stress the fact that all results are valid already at finite $N$!
In section 4, we recast the D(-1)--D3 lagrangians as a topological
0+0 dimensional theory of ADHM constraints and clarify
the connection with the results of \cite{topo}. The fact that
the multi-instanton measure can be related to the partition
function of a topological gauge theory encourage us
to believe that the cumbersome integrations over the multi-instanton
moduli space could be performed directly along
the lines of \cite{nek1}.       
Finally in section 5, we consider the case where the D3-branes wrap
the $\real^4/\zet_p$ singularity. D(-1) instantons are known \cite{dmjm} 
to realize the Kronheimer-Nakajima construction of self-dual
solutions of SYM$_4$ living on a blown down ALE space.
Following the line of our previous analysis we show how the
expected $AdS_5/\zet_p\times S^5$ geometry, with $\zet_p$
acting on the $AdS$ boundary, is recovered from the
probe perspective. In the appendices we have collected some material
concerning the spinorial representation of $SO(10)$ which set up
the notation of section 2, and a rederivation of the
action of section 4 from orbifold projection on the ${\cal N}=4$
topological action of \cite{dhk}.

\section{Instanton moduli spaces from D-branes dynamics}

The starting point of our discussion are the results of \cite{dhkmv} for
the measure and multi-instanton action in the ${\cal N}=4$ case
\footnote{We denote by
${\cal N}$ one quarter of the number of real supercharges.}
which, when suitably modified, will lead to the models of interest in this paper.
The basic objects in the ADHM construction \cite{adhm} of $SU(N)$
instanton solution in four dimensions are the
$[N+2k]\times [2k]$ and $[2k]\times [N+2k]$
matrices
\begin{equation}
\Delta(x)
\ =\ a \ +\
b \, x \ \ ,\quad\quad
\bar{\Delta} (x)= \bar{a}+ \bar{x} \bar{b} \ \ ,
\label{del}
\end{equation}
where $x_{\alpha\dalpha}=x_m \sigma^m_{\alpha\dalpha}$,
$\bar{x}^{\dalpha\alpha}=x^m \bar{\sigma}_m^{\dalpha\alpha}$, 
\footnote{We adopt the following convention for the sigma matrices: 
$\sigma^m=(\sigma^c,-i{\bf 1})$ and $\bar{\sigma}^m=(-\sigma^c,-i{\bf 1})$.}
is the 
position of the multi-instanton center of mass and all the remaining
moduli are collected in the matrix 
$a$ (see formula (\ref{aM}) below).
Finally $b$ is a $[N+2k]\times [2k]$ matrix which
can be conveniently chosen to be
\be  
 b= \pmatrix{0 \cr {\bf 1}_{[2k]\times[2k]}}
\quad\quad
\bar{b}=\big(0 \ ,\ {\bf 1}_{[2k]\times[2k]}
\big) \ \ .
\label{bs}
\ee
The moduli space of the solutions to the self-dual equations
of motion is characterized in terms
of the supercoordinates
\be
a \equiv \pmatrix{  w_{u \, i \dalpha}\cr
a_{\alpha \dalpha} ~_{li}}
\quad
{\cal M}^A \equiv
\pmatrix{ \mu^A_{iu} \cr
{\cal M}^{ A}_\beta ~_{li}}
\label{aM}
\ee
satisfying the bosonic and fermionic ADHM constraints
\bea
\bar{\Delta}\Delta &=& f^{-1}_{k\times k} {\bf 1}_{[2]\times[2]} \nonumber\\
\bar{\Delta}{\cal M}^A &=& \bar{\cal M}^A \Delta \ \ ,
\label{constr}
\eea
with $f_{k\times k}$ an invertible
$[k]\times [k]$ matrix.

The solutions to the 
self-dual equations of motion for the 
various fields in the ${\cal N}=4$ vector multiplet 
are given by \cite{dhkmv}
\bea
A_n &=& 
\bar{U} \ \partial_{n}
\ U \nonumber\\
\Psi^A &=& \bar{U}\left( {\cal M}^A f \bar{b}-b f \bar{{\cal M}}^A\right)U \nonumber\\
i\,\Phi^{AB}(x)  &=&
{1\over2\sqrt{2}}\,\bar{U}\,\big({\cal{M}}^B f\bar{\cal M}^A
-{\cal M}^A f\bar{\cal M}^B   \big)U\ +\nonumber\\
&&+  \bar{U} \cdot
\pmatrix{
0_{[N]\times [N]}& 0_{[N]\times[2k]} \cr
0_{[2k]\times[N]} 
&L^{-1}\Lambda^{AB}_{ [k]\times[k]}\otimes 1_{[2]\times[2]}}
\cdot U 
\label{An}
\eea
in terms of the kernels $U_{[N+2k]\times [N]},~\bar{U}_{[N]\times [N+2k]}$ 
of the ADHM matrices $\bar{\Delta},\Delta$.
In (\ref{An}), $\Lambda^{AB}$ is the fermionic bilinear
\be
\Lambda^{AB}={1\over{2\sqrt{2}}}
\left(\bar{\cal M}^A{\cal M}^B - \bar{\cal M}^B{\cal M}^A\right) \ \ ,
\label{lambda}
\ee
and the operator $L$ is defined as
\be
L\cdot \Omega = {1\over 2} \{W^0,\Omega\} + [a_m,[a_m,\Omega]] \ \ ,
\label{L}
\ee
with $W^0_{ij}=\bar{w}^{\dalpha}_{i u}w_{u j \dalpha}$.
In the above equations the indices $i,j=1,\ldots,k$,  
$u,v=1,\ldots,N$, $\alpha,\dalpha=1,2$ and $A=1,\ldots,4$ label the
instanton number, the color, $SO(4)$ and 
$SO(6)_{\cal R}\cong SU(4)_{\cal R}$ quantum numbers, 
respectively.

The measure of the moduli space and the multi-instanton action
can be efficiently read from the partition
function of the gauge theory governing the
low energy dynamics of a system of $k$ D(-1) and $N$ D3-branes
moving in flat space \cite{dhkmv}
\bea
Z_{k,N} = \frac{1}{{\rm Vol}\,U(k)}\,
\int d^{6 k^2}\chi\, d^{8 k^2}\lambda\, d^{3 k^2}D\,d^{4 k^2}a\,
d^{8 k^2}{\cal M}\,
d^{2 kN}w\,d^{2kN}\bar w\, d^{4kN}\mu\, d^{4kN}\bar{\mu}\,
e^{-S_{k,N}}\ , \nonumber\\
\label{partfunc}
\eea
after integrating out the $(\chi_a,\lambda_{\dalpha A},D_c)$
degrees of freedom associated to the $U(k)$ vector
multiplet, where $a=1,\ldots,6$ stands for the vector index in the 
transverse $SO(6)_{{\cal R}}$ and $c=1,2,3$.
This gauge theory is defined by the dimensional reduction
to $0+0$ dimensions of the ${\cal N}=1$ $U(k)$ gauge theory
in $D=6$ with one hypermultiplet transforming in the adjoint representation,
denoted by $(a_{\alpha\dalpha},{\cal M}_\alpha^A)$
and $N$  transforming in the fundamental representation (and its conjugate),
denoted by $(w_{\dalpha},\mu^A ;\bar{w}^{\dalpha},\bar{\mu}^A)$.
The action can be written as
\be
S_{k,N}={1\over g_0^{2}}S_{G} + S_{K}+S_{D}
\label{cometipare}
\ee
with
\bea
&&S_{G}={\rm tr}_{k}\big(-[\chi_a,\chi_b]^2+\sqrt{2}i\pi
\lambda_{\dot{\alpha} A}[\chi_{AB}^\dagger,\lambda^{\dot{\alpha}}_B]
-D^{c}D^{c}\big)
\label{Sd}\\
&&S_{K}=-{\rm tr}_{k}\big([\chi_a,a_{n}]^2
-\chi_a \bar{w}^{\dalpha}
w_{\dalpha}\chi_a + \sqrt{2}i\pi
{\cal M}^{\alpha A}[\chi_{AB}
{\cal M}^{B}_{\alpha}]-2\sqrt{2} i \pi
\chi_{AB} \bar{\mu}^{A} \mu^{B} \big)\, \nonumber\\
&&S_{D}={\rm tr}_k\,\big(i \pi\left(
-[a_{\alpha\dot{\alpha}},{\cal M}^{\alpha A}]
+\bar{\mu}^{A} w_{\dot{\alpha}}
+\bar{w}_{\dot{\alpha}}\mu^{A}\right)
\lambda^{\dot{\alpha}}_{A}
+D^{c}\left(\bar{w} \tau^c w-i \bar{\eta}_{mn}^c [a_{m},a_{n}]
\right)\big)\nonumber
\eea
Given the classical group isomorphism $SO(6)_{\cal R}\cong SU(4)_{\cal R}$, 
$SO(6)_{\cal R}$
vectors can also be written as
$\chi_{AB}\equiv {1\over \sqrt{8}} \Sigma^a_{AB} \chi_a$ with the
$\Sigma^a_{AB}=(\eta^c_{AB},i\bar\eta^c_{AB}), \bar\Sigma^a_{AB}=
(-\eta^c_{AB},i\bar\eta^c_{AB})$ given in terms of the t'Hooft symbols.
In the limit $g_0\rightarrow\infty$, gravity decouples from
the gauge theory and the contributions coming from $S_G$
are suppressed; then the fields $\lambda_{\dalpha}^A$, $D_c$
become lagrangian multipliers implementing the ADHM constraints
(\ref{constr}) in the form
\bea
&&\bar{\mu}^{A} w_{\dot{\alpha}}
+\bar{w}_{\dalpha}\mu^{A}
-[a_{\alpha\dot{\alpha}},{\cal M}^{\prime\alpha A}]=0\nonumber\\
&& \bar{w} \tau^c w-i \bar{\eta}_{mn}^c [a_{m},a_{n}]=0
\label{adhm}
\eea
An explicit evaluation of the remaining integrations
for generic $k$ seems to be at present out of reach.
However, in the limit of large $N$ significant
simplifications take place. This program has been carried out in
\cite{dhkmv}. The results show that the degrees of freedom associated
to the $k$-instanton center of mass are described by a position in
an $AdS_5\times S^5$ space, while the dynamics of the
excitations around this configuration are governed by a $0+0$
gauge theory obtained from the dimensional reduction of a ${\cal N}=1$
$SU(k)$ gauge theory in $D=10$ to zero space-time dimensions.

In the following we apply these techniques to
non conformal gauge theories and show how to make contact with
the geometry of the corresponding supergravity backgrounds already
at finite $N$.

\section{On the geometry of the non conformal case}

Having seen how the ADHM moduli space for ${\cal N}=4$ arises from a 
D(-1)--D3 brane system 
\footnote{Our choice for the dimensionality of the branes 
is taken for the sake of clarity and will not spoil the generality of 
our discussion.}
, this section is devoted to the study of gauge theories living on D3-branes
located at a $\real^4/\Gamma$ orbifold singularity, $\Gamma$ being 
a discrete subgroup of $SU(2)$.
%By a simple scaling analysis we isolate that part of
%the measure which represents the geometry of the near horizon.
%The rest of the measure is cast in a form which is not amenable to a simple
%treatment but can be seen not to distort the geometry left aside the running
%of the coupling constant.

%The generalization to $\real^6/\Gamma$ with
%$\Gamma=\zet_p$\footnote{For definiteness we
%choose branes in the identity representation of $\Gamma$.
%We will deal with more general situations later on.} yields to cases with
%various supersymmetries ${\cal N}=0, 1, 2$ and matter content.

\subsection{Pure ${\cal N}=2$ Super Yang-Mills}

We first discuss the case of pure ${\cal N}=2$ $SU(N)$ 
SYM$_4$ to exemplify
our strategy, postponing to the next subsection the most general case
corresponding to gauge theories with product groups and bifundamental
matter. 

In the following our D3-branes will be longitudinal to the 7, 8, 9, 10
directions of space time and transverse to the remaining ones. In this
section discrete symmetries act on the plane formed by the 1, 2, 4, 5 
directions leaving 3, 6 unaltered. 
The low energy dynamics for a stack of $N$ fractional D3-branes lying at 
a $\real^4/\Gamma$ singularity 
is governed by a pure ${\cal N}=2$ 
$SU(N)$ SYM$_4$. This theory can be obtained
by means of a $\Gamma$-projection of the previously discussed 
${\cal N}=4$ $SU(N)$ 
gauge theory associated to $N$ nearby D3-branes moving  
on a flat space-time. In the ${\cal N}=2$ language, fields in 
the vector multiplet of the parent ${\cal N}=4$ 
theory are invariant under $\Gamma$, while the
whole hypermultiplet, whose fields roughly describe  
the positions of the branes
in the orbifold space, is projected out. 
Fractional D3-branes are then stuck at the origin of the orbifold space.
Alternatively one can think of these branes 
as D5-branes wrapping a vanishing 
(at the orbifold point) two-cycle of a blown down ALE space. 

The aim of this section is to show how the moduli space measure and
multi-instanton action of pure ${\cal N}=2$ $SU(N)$ Super Yang-Mills theory 
can be read from the low energy lagrangian governing the
dynamics of a fractional D(-1)--D3 system on $\real^4/\zet_p$.
The low energy excitations of such a system are defined by
the $\Gamma$-projection \cite{dmjm}, which in the notation of \cite{hk}
reads
\bea
\mu^A &=& e^{2\pi i {q_A \over p}} \mu^A \quad\quad
\bar{\mu}^A = e^{2\pi i {q_A \over p}}
\bar{\mu}^A \nonumber\\
{\cal M}^A_{\alpha} &=& e^{2\pi i {q_A \over p}}
{\cal M}^A_{\alpha} \quad\quad
{\lambda}^A_{\dalpha} = e^{2\pi i {q_A \over p}}
{\lambda}^A_{\dalpha} \nonumber\\
\chi^{AB} &=& e^{2\pi i {q_A+q_B \over p}} \chi^{AB}
\label{Ginv}
\eea
with $q_1=q_2=0$, $q_3=-q_4=1$. All the other fields 
$(w_{\dalpha},\bar{w}^{\dalpha},a_m,D^c)$
remain invariant under the $\zet_p$ orbifold action.

The net effect of the projection (\ref{Ginv}) is to break the $SU(4)$ 
${\cal R}$-symmetry group of the ${\cal N}=4$ theory down to 
$SU(2)_{\dot{A}}\times U(1)_{\cal R}$,
where $SU(2)_{\dot A}$ is the ${\cal N}=2$ automorphism 
group and $U(1)_{\cal R}$ the anomalous ${\cal R}$ charge. A more detailed 
discussion  of this point can be found in the appendix A.

After the $\Gamma$-projection (\ref{Ginv}) the multi-instanton
action can be read from (\ref{Sd}) with fermionic
indices $A,B$ now restricted to $\dot{A},\dot{B}=1,2$ 
(in the fundamental of the automorphism group)
and the
bosonic components $(\chi_1, \chi_2, \chi_4, \chi_5)$ set to 
zero.
The remaining components will be denoted by the complex notation
\bea
\phi &\equiv& 2\chi_{43}=2\chi^{12}={1\over \sqrt{2}} (-\chi_3+i \chi_6)
\nonumber\\
\bar{\phi} &\equiv& 2\chi_{21}=2\chi^{34}={1\over \sqrt{2}} (-\chi_3-i \chi_6) \ \ .
\label{phi}
\eea

We can now follow \cite{dhkmv,hkm} in order to determine a 
gauge invariant measure. We start by defining the 
$SU(N)$ gauge invariant components $W^m_{ij},\zeta^{\dalpha \dot{A}}_{ij}$
through
\bea
W^0_{ij} &=&\bar{w}_{iu}^{\dalpha}
 \,w_{uj\dalpha}\ ,\quad \quad\quad\quad\quad W^c_{ij}=\bar{w}_{iu}^{\dalpha}
\, \tau^c {}^{\dbeta}_{\dalpha} \,w_{uj\dbeta}~~~ c=1,2,3\nonumber\\
\mu_{iu}^{\dot{A}} &=& w_{uj\dalpha}\zeta^{\dalpha \dot{A}}_{ji}+
\nu_{iu}^{\dot{A}},\quad\quad\quad
{\rm with}~~\bar{w}_{iu}^{\dalpha} \nu_{uj}^{\dot{A}}=0, \nonumber\\
\bar{\mu}_{iu}^{\dot{A}} &=& \bar{\zeta}^{\dot{A}}_{\dalpha ij}
\bar{w}_{uj}^{\dalpha}
+\bar{\nu}_{iu}^{\dot{A}}\quad\quad\quad\quad
 {\rm with}~~\bar{\nu}_{iu}^{\dot{A}}
 w^{}_{uj\dalpha}=0\
\label{invv}
\eea
and perform the integrations over the iso-orientation
modes (parameterizing a point in the
coset space ${SU(N) \over SU(N-2k)\times U(1)}$) and over their
``fermionic superpartners'' $\nu^{\dot{A}}$, $\bar{\nu}^{\dot{A}}$
\bea
&&\int_{\rm gauge\atop coset}d^{2kN}w\, d^{2kN}\bar w
=c^{\prime}_{k,N}\left({\rm
det}_{2k}W\right)^{N-2k} \, d^{k^2}W^0 d^{3k^2}W^c\, 
\label{cose}\\
&&\int d^{2k(N-2k)}\nu \, d^{2k(N-2k)}\bar{\nu} \,
{\rm exp} \big[\sqrt8\pi
i{\rm tr}_k
\,\chi_{\dot{A}\dot{B}}\bar{\nu}^{\dot{A}} \nu^{\dot{B}}\big]=
(-2\pi^2)^{k(N-2k)}
\left({\rm det}_{k}\bar{\phi}\right)^{2(N-2k)}\nonumber
\eea
with
\be
c^{\prime}_{k,N}=
{2^{2kN-4k^2+k}\pi^{2kN-2k^2+k}\over\prod_{i=1}^{2k}(N-i)!} \ \ .
\label{ckn}
\ee
Furthermore, by integrating out the auxiliary 
field $D^c$ we enforce the bosonic ADHM constraint $W^c=i\bar{\eta}^c_{mn}[a_n,a_m]$,
while the integration on $\lambda_{\dalpha}^{\dA}$ enforces the fermionic 
constraint and give rise to a factor $\pi^{4k^2}$ in the partition function.
Plugging (\ref{invv}) and (\ref{cose}) into the $\zet_p$ projected analog of
(\ref{partfunc}) one is left with the $SU(N)$ invariant measure \cite{hkm}
\bea
Z_{k,N} &=& c_{k,N} 
e^{2\pi i k \tau}
 {1 \over {\rm Vol}\, U(k)} \int d^{k^2} W^0\, d^{4k^2}a \,
d^{k^2}\phi\,d^{k^2}\bar{\phi}\,
d^{4k^2}{\cal M} \, d^{4k^2}\zeta \,
\nonumber\\
&& \times \left({{\rm det}_{2k} W} {\rm det}_{k}\bar{\phi}^2\right)^{N-2k}\,
e^{-S_{k,N}} \ \ ,
\label{zkn2}
\eea
with
\bea
S_{k,N} &=& - 4\pi i \,{\rm tr}_k\,\bar{\phi}
{\Lambda}^{12}+ {\rm tr}_k\,(\bar{\phi}L \phi+ \phi L \bar{\phi}) \, \label{l2}\\
{\Lambda}^{\dot{A}\dot{B}}&=&{1\over{2\sqrt{2}}}
\left(\zeta^{[\dot{A}} W \zeta^{\dot{B}]}
%-\zeta^{\dot{B}} W \zeta^{\dot{A}}
+[a^{\alpha \dalpha}, {\cal M}_{\alpha}^{[\dot{A}}]\zeta_{\dalpha}^{\dot{B}]}
%-[a_{\alpha \dalpha}, {\cal M}^{\alpha \dot{B}}]\zeta^{\dalpha \dot{A}}
+{\cal M}^{\alpha [\dot{A}}{\cal M}_{\alpha}^{\dot{B}]}\right) \ \ ,
\nonumber
\eea
and
$c_{k,N}=c^{\prime}_{k,N}(-2\pi^2)^{k(N-2k)}\pi^{4k^2}$.
%and $W^{\dalpha}_{\dbeta}={1\over 2} W^m (t^m) ^{\dalpha}_{\dbeta}$,
%with $W^c$ given by the ADHM constraint $W^c=i\bar{\eta}^c_{mn}[a_n,a_m]$.
Here and in the following we denote by 
$\Phi^{[AB]}\equiv \Phi^{AB}-\Phi^{BA}$ the antisymmetrization in the 
$SU(4)_{\cal R}$ indices.

We would like now to show that all dependence on the center of mass
degrees of freedom in (\ref{zkn2}) factorizes in an $AdS_5\times S^1$
term already at finite $N$. Inspired by large $N$ manipulations we
split each $U(k)$ adjoint V-field into its trace and traceless part
\be
V= v_0 {\bf 1}_{k\times k} + \hat{v} \ \ .
\ee
At this point it is convenient to rescale the various
traceless components of the fields in the following
way
\bea
a_m &=& -x_m+ \rho \hat{a}_m
\nonumber\\
\phi &=& r e^{i \vartheta}\left(1+\hat{\phi}\right)
%      =  r e^{i \vartheta}\tilde{\phi} 
\nonumber\\
\bar{\phi} &=& r e^{-i \vartheta}\left(1+\hat{\bar{\phi}}\right)
%            =  r e^{-i \vartheta}\tilde{\bar{\phi}}
\nonumber\\
W_0 &=& \rho^2 \left(1+\hat{W}_0\right)
%     = \rho^2 \tilde{W}_0
\nonumber\\
\zeta_{\dalpha}^{\dot{A}} &=& \bar{\eta}_{\dalpha}^{\dot{A}}
+\rho^{-{1\over 2}} e^{i\vartheta \over 2}
\hat{\zeta}_{\dalpha}^{\dot{A}}\nonumber\\
{\cal M}_{\alpha}^{\dot{A}} &=& {\xi}_{\alpha}^{\dot{A}}
+ \rho \hat{a}_{\alpha \dalpha}\bar{\eta}^{\dalpha \dot{A}} +
\rho^{{1\over 2}} e^{i\vartheta \over 2}\hat{\cal{M}}_{\alpha}^{\dot{A}} \ \ .
\label{splitting}
\eea
From (\ref{splitting}) it follows that the variables $x_m$, $\rho^2$,
${\xi}_{\dalpha}^{\dA}$, $\bar{\eta}_{\dalpha}^{\dot{A}}$ are
the trace components associated
to the overall bosonic and fermionic translational/conformal zero modes,
while $r$ is the distance of the D-instanton probe from the D3-branes
\footnote{Actually $r=d/\alpha^{\prime}$, where $d$ is the distance from
the D3-branes in the orbifold fixed plane. 
%To simplify the notation,
%in the rest of the paper we will always mean $r$ to be measured in units of 
%$\alpha^\prime$
}.
Our results can be expressed in terms of this quantity and of 
the adimensional variable $y=r\rho$.
In fact, plugging (\ref{splitting}) in (\ref{zkn2}) and taking care
of the Jacobian involved in the rescalings, one can easily check that
all dependence in the center of mass variables
$r, x_m, \vartheta$ factors out in a
``deformed'' $AdS_5\times S^1$ factor
\bea
Z_{k,N} &=& d_{k,N} \int
d^{4}x_m \, r^3 dr d\vartheta \,
e^{4i\vartheta} \, e^{-2 i k N\vartheta} r^{-2kN}
e^{2\pi i k \tau_0} \,
d^{4}{\xi} \, d^{4}\bar{\eta}
\nonumber\\
&=&d_{k,N} \int 
d^{4}x_m \, r^3 dr d\vartheta \,
e^{4i\vartheta} \, e^{-2 i k N \vartheta}
e^{2\pi i k \tau(r)} \,
d^{4}{\xi} \, d^{4}\bar{\eta} \ \ .
\label{partfuncn=2}
\eea
Remarkably the entire deviation from the $AdS_5\times S^1$
measure,  which, up to numerical coefficients, is given by the factor
$d^4 x_m \, r^3dr d\vartheta$ in (\ref{partfuncn=2})
\footnote{We use coordinates \cite{Maldacena:1998re}
where the $AdS_5\times S_1$ metric
reads $ds^2={r^2\over R^2} d x_m^2+{R^2 \over r^2} dr^2 +R^2 d\vartheta^2$
with $R^2=\sqrt{4\pi g_s N}$, $g_s$ being the string coupling constant.},
is reabsorbed in the running coupling constant
\be
2\pi i \tau(r)\equiv 2\pi i \theta-{8\pi^2 \over g(r)^2}
=2\pi i \theta-{8\pi^2\over g_0^2}- 2N \, {\rm ln} (r)=
2\pi i\tau_0 - 2N \, {\rm ln} (r) \ \ .
\label{pi}
\ee
In (\ref{pi}) we identify
the AdS radial coordinate $r$ with the dynamical energy scale $\mu$ 
in the SYM$_4$ gauge theory and read from the term multiplying the logarithm
the first coefficient in the expansion of the $\beta$ function 
\be
\beta\equiv r {d \over dr} g(r)=- b_1 g^3/16\pi^2+\ldots \ \ .
\label{beta}
\ee 
with $b_1=2 N_c-N_F=2N$, where $N_c$, $N_F$  are the number of colors 
and of the fundamental hypermultiplets respectively.

Throughout this paper we employ units for which $\alpha^\prime=1$ and
choose to work with the dimensioned instanton measure $Z_{k,N}$
rather than with the physical measure. The two differ by a numerical
coefficient (see \cite{dhkmv}) and a cut-off $\Lambda$ dependent factor.
Bringing back dimensions in (\ref{cometipare}), (\ref{Sd}) (recalling that
$g_0\sim (\alpha^{\prime})^{-1}$) one can easily check that 
the dimensionless combination is precisely $\Lambda^{2kN} Z_{k,N}$.
The appearance of the cut-off dependent factor in the normalized
instanton measure can be rigorously derived from a Pauli-Villars 
regularization of the fluctuation determinants around 
the instanton background,
see \cite{nsvzmrs} for a complete discussion on this point.
The net result of this normalization is to replace $r$ in (\ref{pi})
by $r/\Lambda$, leading to the expected formula for the running 
coupling constant. 

The extra phase factor $ e^{-2 i (k N - 2) \vartheta}$ in (\ref{partfuncn=2})
implements the selection rule imposed by the presence of the
chiral anomaly.
Indeed, taking into account that each insertion of an
extra fermion, besides the eight associated to conformal and
translational zero modes, carries a power of $e^{i \vartheta \over 2}$,
one needs precisely $4(k N - 2)$ insertions 
in order to cancel out the extra phase factor. This means that
a non-trivial correlation function should involve a total
number of $n=4 k N$ fermionic insertions, as the usual selection
rule of pure ${\cal N}=2$ SYM requires. 
For example, it would be interesting to consider the insertion of single 
or multi-trace composite operators of scalar fields, which have been recently 
shown not to get perturbative anomalous dimensions in 
${\cal N}=2$ theories \cite{Maggiore:2001}.

Finally, the $d_{k,N}$ coefficients are given by the $SU(k)$ integrals
\bea
d_{k,N} &=& c_{k,N}
%% {(8\pi^2)^{k(N-2k)}}
{1 \over {\rm Vol}\, SU(k)} \int d \hat{W}^0\, d \hat{a} \,
d\hat{\phi}\,d\hat{\bar{\phi}}\,
d\hat{{\cal M}} \, d\hat{\zeta} \,dy \nonumber\\
&&\times y^{4 k N - 2 k^2 - 5}
\left(det_{2k} \widetilde{W}
det_{k}\tilde{\bar{\phi}}^2 \right)^{N-2k}\, e^{-S(y)} \ \ ,
\label{dkn}
\eea
where we have introduced the compact tilded notation for the fields
$\widetilde{W}=1+\hat{W}$ and $\tilde{\phi}=1+\hat{\phi}$. The action appearing
in (\ref{dkn}) reads
\be
S(y)=-4\pi i y\,{\rm tr}_k\, \tilde{\bar{\phi}}
\tilde{\Lambda}^{12}-y^2 {\rm tr}_k\, 
(\tilde{\phi}\tilde{L} \tilde{\bar{\phi}}+
\tilde{\bar{\phi}}\tilde{L} \tilde{\phi}) \ \ , 
\ee
with $\tilde{\Lambda}^{12}$ given by the same expression (\ref{l2})
as before but now in terms of the $SU(k)$ components 
of the fermionic zero-modes 
and of the tilded field $\widetilde{W}$.
For our purposes it will not be important to compute 
these coefficients,
since they do not involve powers of $r$ and therefore 
will not affect the associated geometry
in (\ref{partfuncn=2}).

\subsection{Adding regular branes}

In this subsection we extend the previous results to 
configurations involving both fractional and regular
D3-branes moving on $\real^4/\zet_p$. 
More general orbifold projections $\real^6/\zet_p$ \cite{kasi}
are specified by four integers
$q_A$ such that $q_1+q_2+q_3+q_4=0~~{\rm mod}~p$ and lead to ${\cal N}=1,0$
supersymmetric gauge theories depending on whether $\zet_p$ is embedded in
an $SU(3)$ or $SU(4)$ subgroup of $SO(6)_{\cal R}\cong SU(4)_{\cal R}$ respectively.
The former case corresponds to 
$q_A\ne 0$ for three values of the index $A$, while the latter to  
$q_A\ne 0$ for all values of $A$.
Fractional instanton 
in these less supersymmetric theories are
constrained to live on the boundary of the $AdS_5$ space 
\footnote{In the previous section and all the time along this paper 
we studied the case 
$\rreal^6/\zzet_p=\rreal^4/\zzet_p\times \rreal^2$ in which fractional 
instantons are allowed to move
in the directions 3,6 of the transverse space. Here we refer
to the case where the transverse space 
is  $\rreal^6/\zzet_p$ and fractional instantons live at the fixed point 
$r=0$.}
and therefore test only the less interesting four-dimensional flat geometry.
This is the main reason why we restrict our analysis to the
${\cal N}=2$ case even if most of our manipulations
apply in the ${\cal N}=0,1$ contexts.
We will follow
closely \cite{hk}, but focusing as in the previous section 
on non conformal brane configurations.
We denote by $N_q$ with $q=0,1,..,p-1$ the number of D3-branes
transforming in the $q^{th}$-representation of $\zet_p$. A regular
brane is defined by a set of $p$ D3-branes, one of each type, and
therefore the general configuration can be thought as built
out of $n={\rm min}_q \, \{ N_q \}$ regular branes
plus $N_q-n$ fractional branes of the $q$-type. The example of
pure ${\cal N}=2$ studied in the previous section corresponds
to the case of no regular branes $n=N_q=0$ for $q>0$
with $N_0=N$ fractional branes of the $0^{th}$-type.
In a similar way we denote by $k_q$ the number of $D(-1)$
instantons of a given type.
We consider first the case where the 
D3-brane geometry is probed by fractional instantons
( $k_q$ generic).
The modifications to the 
case where regular instantons $k_0=k_2=....=k_{p-1}=K/p$
are used as test charges will be commented only at the end of this 
section.  

The analysis of the resulting gauge theory follows closely
the one in the previous section but taking care of the
fact that now
the $\zet_p$ projection acts on both Chan-Paton and $SU(4)_{\cal R}$ indices.
The precise action on the various D(-1)--D3 fields is given by
\bea
w_{\dalpha} &=&\gamma_N \, w_{\dalpha} \, \gamma_K^{-1}
\quad\quad 
\mu^A = e^{2\pi i {q_A \over p}} \gamma_N \, \mu^A
\,\gamma_K^{-1} \nonumber\\
\bar{w}^{\dalpha} &=&\gamma_K \, \bar{w}^{\dalpha} \, \gamma_N^{-1}
\quad\quad
\bar{\mu}^A = e^{2\pi i {q_A \over p}} \gamma_K \, \bar{\mu}^A
\,\gamma_N^{-1} \nonumber\\
a_{\alpha\dalpha} &=&
\gamma_K \, a_{\alpha\dalpha}\, \gamma_K^{-1}\quad\quad
{\cal M}^A_{\alpha} = e^{2\pi i {q_A \over p}}
\gamma_K \, {\cal M}^A_{\alpha}\, \gamma_K^{-1}\nonumber\\
\chi_{AB} &=& e^{2\pi i {q_A+q_B \over p}}
\gamma_K \, \chi_{AB}\, \gamma_K^{-1}\quad
D^c = \gamma_K \, D^c\, \gamma_K^{-1}\quad
{\lambda}^A_{\dalpha}= e^{2\pi i {q_A \over p}}
\gamma_K \, {\lambda}^A_{\dalpha} \, \gamma_K^{-1}
\nonumber
\eea
where $q_1=q_2=0, q_3=-q_4=1$, and
$\gamma_K$, $\gamma_N$ are $K\times K$ and $N\times N$ matrices
(with $N=\sum_q N_q$ and  $K=\sum_q k_q$) realizing the orbifold
group action on the Chan-Paton indices.
The surviving components can be written as
\bea
w_{\dalpha}^q &=& w_{\dalpha u_q i_q}\quad \quad\quad
\mu^A_{q} = \mu^A_{u_q i_{q+q_A}} \nonumber\\
\bar{w}^{\dalpha}_q &=& \bar{w}_{\dalpha i_q u_q}\quad\quad\quad
\bar{\mu}^A_{q} = \bar{\mu}^A_{i_{q+q_A} u_q}
\nonumber\\
a^{\alpha\dalpha}_q &=& a^{\alpha\dalpha}_{i_q j_q}\quad\quad~~~~
{\cal M}^{A\alpha}_q = {\cal M}^{A\alpha}_{i_{q+q_A} j_q} \nonumber\\
\chi^{AB}_q &=& \chi^{AB}_{i_{q+q_A} j_{q-q_B}} \quad\quad
D^c_q= D^c_{i_q j_q}\quad\quad
\lambda^{A\dalpha}_q = \lambda^{A\dalpha}_{i_{q+q_A} j_q} 
\label{surv}
\eea
where $q= 0,1,...,p-1$, $i_q=1,...,k_q$, 
$u_q=1,...,N_q$ are the block indices\footnote{More precisely the 
indices in (\ref{surv}) run as 
$i_1=1,..,k_1, i_2=k_1+1,..,k_1+k_2$
and so on. We adopt the simplifying 
notation where the subscripts ``$q$'' indicates the 
matrix blocks from which $i_q,u_q$ start to count.}. 
All sums in $q$ are from now on understood modulo $p$.

The result (\ref{surv}) is derived from an iterative application
of the Mac-Kay correspondence.
We start by defining $K$($N$)-dimensional vector spaces 
$V$($W$). 
Under the action of $\Gamma=\zet_p$ they decompose as $V= \sum_q k_q V_q$
($W=\sum_q N_q W_q$)  where $\{ k_q \}$($\{ N_q \}$) labels the number of
D(-1) (D3)-branes transforming in the $q^{th}$-irreducible 
representation $R_q$ of $\zet_p$ and
$V_q$($W_q$) are one-dimensional vector spaces associated to this
irreducible representation.
Open string modes can be thought as homomorphisms (or endomorphisms)
between the real vector spaces $V$, $W$. In addition to the above
decomposition in the Chan-Paton space, open string modes carrying
a $\bar{A}=3,4$ index transform in the so called regular defining 
representation $Q$ of $\zet_p$.
The Mac-Kay correspondence states that the representation 
$R_q\otimes Q$ decompose under the action of a discrete
subgroup $\Gamma$ of $SU(2)$ as
\be
 R_q\otimes Q=\oplus_r A_{qr} R_r
\label{mckay}
\ee
with 
$A_{qr}=2\delta_{qr}-\tilde{C}_{qr}$, $\tilde{C}_{qr}$
being 
the extended Cartan matrix of the A-D-E algebra associated
to $\Gamma$. In the present situation 
$\Gamma=\zet_p$ and $A_{q r}=\delta_{q, r+1}+\delta_{q, r-1}$. 
Collecting together $\dot{A}=1,2$ and $\bar{A}=3,4$ 
indices in the ``$A$'' index, we can write (\ref{mckay}) as 
$R_q\otimes Q_{A}=\delta_{q,q+q_A} R_q$ with $A=1,..,4$.
The $\Gamma$ invariant components can then be easily found using 
the Schur' s lemma $Hom (R_q, R_r)_{\Gamma}=\delta_{qr}$.
Let us illustrate how this works 
for the fields ($\chi_{AB},\lambda^A_{\dalpha},D^c$)
in the vector multiplet:
\bea
\chi_{AB}:&&~~~{\rm Hom}\, ( Q_A \otimes V,\bar{Q}_B\otimes V)_{\Gamma}=
{\rm Hom}\,({R}^{k_{q+q_A}},{R}^{k_{q-q_B}})\nonumber\\
\lambda^A_{\dalpha}:&&~~~          
 {\rm Hom}_{\dalpha}\,( Q_A \otimes V, V)_{\Gamma}=
{\rm Hom}_{\dalpha}\,({R}^{k_{q+q_A}},{R}^{k_{q}})\nonumber\\
D^c:&&~~~{\rm Hom}_{c}\,(V, V)_{\Gamma}=
{\rm Hom}_{c}\,({R}^{k_{q}},{R}^{k_{q}}) \ \ .
\eea
It is a straightforward exercise to extend this analysis to the remaining
fields of the D(-1)--D3 lagrangian, 
leading finally to the matrix block form (\ref{surv}).

Coming back to (\ref{surv}) one notice that after the $\Gamma$
projection the D(-1) gauge group is reduced to $\prod_q U(k_q)$ 
\footnote{The existence
of this ''auxiliary'' $\prod_q U(k_q)$ corresponds to 
the internal symmetries of the ADHM construction. See 
\cite{topo,Bruzzo:2000di} for a more detailed discussion of its role.}. 
%From (\ref{surv}) one can see that the spectrum of the low-energy
%excitations of this system 
%consists of gauge bosons $(k_q,\bar{k}_q)$ in the product gauge group 
%along with $N_q$ matter fields that transform in the fundamental
%representation (and its conjugate) of the $U(k_{q+q_A})$ group factor. 
A similar projection on the D3--D3 strings shows
that the four-dimensional gauge group is reduced to
$\prod_q U(N_q)$, and that the field theory spectrum 
is given by vector multiplets in the adjoint representation 
of the product gauge group and matter hypermultiplets 
in the bifundamental representations $(\bar{N}_q, N_{q+q_A})$.
 
{\it Mutatis mutandis} we follow the strategy of the previous subsection,
see (\ref{invv})--(\ref{ckn}). 
The partition function (\ref{zkn2}) generalizes to
\bea
{\cal Z}_{k, N} &=& \prod_q \Big\{ c_{k_q,N_q} e^{2\pi i k_q \tau_q}
 {1 \over {\rm Vol}\, U(k_q)} \int d^{k_q^2} W^0\, d^{4k_q^2}a \,
d^{2 k_q (4\hat{k}_q - k_q)}\chi \,
d^{8 k_q \hat{k}_q}{\cal M} \, d^{8 k_q \hat{k}_q}\zeta \,
\nonumber\\
&& \times \left({{\rm det}_{2k_q} W} {\rm det}_{4 \hat{k}_q}
\chi \right)^{N_q-2k_q} \Big\} \,
\exp\big[4\pi i\,{\rm tr}_k\,\chi_{AB}
{\Lambda}^{AB}- {\rm tr}_k\, \chi_a L \chi_a \big]\, .
\label{zkn4}
\eea
with $\hat{k}_q={1\over 4}\sum_{A=1}^4 k_{q+q_A}$.

%Since we are considering ${\cal N}=2$ orbifold group projections,
%we can restrict ourselves without loss of generality to
%$q_A=(0,0,1,-1)$.
The  $\dA,\dB=1,2$ components of the fields in (\ref{surv})
can be split exactly as in the previous formula (\ref{splitting}) into their
trace and traceless part,  
where now the instanton center of mass degrees of freedom clearly belong 
to the block diagonal components in (\ref{surv}),
while the $SU(k_q)$ fields are given in terms of 
still block diagonal  
but traceless $q\times q$ matrices.

The remaining $\bar{A},\bar{B}=3,4$ components in (\ref{surv}) are
instead off-diagonal, and can be conveniently rescaled as  
\bea
\phi^s &=& r e^{-i \vartheta} \hat{\phi}^s
%\tilde{\phi}^s
\nonumber\\
\bar{\phi}^s &=& r e^{i \vartheta} \hat{\bar{\phi}}^s
%\tilde{\bar{\phi}}^s
\nonumber\\
\zeta_{\dalpha}^{\bar{A}} &=&
 \rho^{-{1\over 2}} e^{-i{\vartheta \over 2}}
\hat{\zeta}_{\dalpha}^{\bar{A}}
%\tilde{\zeta}_{\dalpha}^{\bar{A}}
\nonumber\\
{\cal M}_{\alpha}^{\bar{A}} &=&
\rho^{{1\over 2}} e^{-i{\vartheta \over 2}}
\hat{\cal{M}}_{\alpha}^{\bar{A}}  \ \ ,
\label{splitting1}
\eea
where by 
\be
\phi^s={1\over \sqrt{2}}(-\chi_s+i\chi_{3+s}) \ \ , ~~s=1,2 
\label{phis}
\ee 
we indicate the two complex coordinates
which span the $\real^4/\zet_p$ space orthogonal both to the D3-brane
and to the orbifold fixed plane.

Plugging in (\ref{zkn4}) one is left with the final result
\footnote{The sum over repeated $q$ indices is always understood.}
\bea
{\cal Z}_{K,N} &=& D_{K,N} \int 
d^{4}x_m \, r^3 dr d\vartheta
\, e^{4i\vartheta} \,
(r e^{i \vartheta})^{-4 N_q (k_q-\hat{k}_q)}
e^{2\pi i \,k_q \tau^0_q }\,
d^{4}{\xi} \, d^{4}\bar{\eta}
\nonumber\\
&=&D_{K,N} \int 
d^{4}x_m \, r^3 dr d\vartheta
\, e^{4i\vartheta} \,
e^{-4 i\vartheta N_q (k_q-\hat{k}_q)}
e^{2\pi i k_q \tau_q(r)}\,
d^{4}{\xi} \, d^{4}\bar{\eta} \ \ ,
\label{zknf}
\eea
where $D_{K,N}$ is a numerical coefficient  
given again in terms of integrals over the $SU(k_q)$ fields  
in (\ref{zkn4}).

We can see from (\ref{zknf}) that the expected 
$AdS_5\times S^1$ volume form corresponding to the multi-instanton 
center of mass degrees of freedom factors out while the additional
powers of $r$ are reabsorbed in the
gauge coupling constants $\tau_q(r)$ associated to the $q$-th gauge group
\bea
2\pi i k_q \tau_q(r)&\equiv& 2\pi i k_q
\theta_q -{8\pi^2 k_q \over g^2_q(r)}\nonumber\\
&=& 2\pi i k_q \theta_q-{8\pi^2 k_q\over g_q^2}-
4N_q (k_q-\hat{k}_q)  \, {\rm ln} (r) \ \ .
\label{running4}
\eea
We can now check (\ref{running4}) against the running of the coupling
constants computed with field theory methods.
As anticipated, the field content of our SYM$_4$ can be
organized in terms of ${\cal N}=2$ supermultiplets as
\bea
{\rm Vector~ multiplets:} &&~~~~~ {\rm Adjoint~~of~~}\prod_q U(N_q)
\nonumber\\
{\rm Hypermultiplets}: &&~~~~~ (\bar{N_q}, N_{q+1})+(\bar{N_q}, N_{q-1})  \ \ ,
\label{multi}
\eea
For each $U(N_q)$ factor we then have one vector multiplet and
$N_F^{(q)}=N_{q+1}+N_{q-1}$ fundamental hypermultiplets.
The first coefficient in the expansion of the beta function 
is given by
\be
b^{(q)}_1=2N_q-N^{(q)}_F=
2N_q-N_{q+1}-N_{q-1}\ \ .
\label{betafunc}
\ee
%\be
%b_q= 2 N^{(q)}_F C({\cal R}_H) - 2 C({\cal R}_V) =
%N_{q+1}+N_{q-1}-2N_q \ \ ,
%\label{betafunc}
%\ee
%$C({\cal R}_{V,H})$ being the index of the representation 
%${\cal R}_{V,H}$
%under which the $V$-vector or $H$-hyper multiplet transforms.
The expression (\ref{betafunc}) agrees with 
the D-instanton result (\ref{running4}) noticing
that
\be
 k_q b^{(q)}_1= k_q (2 N_q - N_{q+1}-N_{q-1})=
4 N_q (k_q -\hat{k}_q)  \ \ .
\label{kb}
\ee
We remark that, as for the pure ${\cal N}=2$ case of the previous 
section, we do not need an explicit expression for the $D_{K,N}$
coefficients in (\ref{zknf}) for our present purposes, since they 
do not affect the resulting geometry. Nonetheless, by studying 
their behaviour in the large $N$ limit, which is amenable to a saddle
point expansion, we can make contact with the
more quantitative results of \cite{hkm}. 
This corresponds to the case 
$k_0=K, k_1=...=k_{p-1}=0$ and $N_0=N_1=...=N_{p-1}=N$, which,
accordingly to (\ref{multi}), 
describe charge $K$ instantons in a ${\cal N}=2$ $U(N)$ theory
with $N_F=2N$ fundamental hypermultiplets \cite{hkm}.

The relevant large $N$ effective action for the bosonic fields can be obtained
along the same lines of \cite{hkm}. In particular, it can be shown that
the dominant saddle point contribution in the large $N$ limit
is the maximally degenerated solution, which corresponds to vanishing 
hatted fields in (\ref{splitting}) and (\ref{splitting1}).
Plugging this solution in the multi-instanton action and exponentiating the
powers of $y$ which appear in the coefficient $D_{K,N}$, one is left with 
the effective action
\be
S_{eff}\sim 2KN( y^2 - 2 {\rm ln}\, y) \ \ ,
\ee
with a saddle point at $y=1$, {\it i.e.}  $r=\rho^{-1}$ \cite{dhkmv,hkm}.

The $SU(K)$ fluctuations around this solution can be  
treated in the large $N$ limit as in \cite{hkm}; in particular,
it turns out that their dynamics is governed by a $0+0$ dimensional
gauge theory obtained from the dimensional reduction of 
a ${\cal N}=1$ $SU(K)$ gauge theory in $D=6$.
Thus in this case the large $N$ limit of the $D_{K,N}$ coefficients
is proportional to 
the partition function $\hat{\cal Z}_K^{(6)}$ of the
above mentioned $0+0$ dimensional gauge theory.

Finally let us consider the case where the D3-brane geometry
is probed by regular D-instantons $k_0=k_1=...=k_{p-1}=K/p$.
First of all notice that while fractional instantons were obliged to live
in the orbifold plane $\phi_s=0$ 
\footnote{We recall that this is the $\real^2$ part of the 
transverse space not acted upon by $\zet_p$.},
regular D-instantons come together with their images
and can roam in the entire transverse space $\real^6$.
Then, the ansatz (\ref{splitting1}) should be 
improved 
in order to take into account the new bosonic and fermionic zero
modes associated to the translational modes along the orbifolded
directions. The $r$ dependence of the instanton measure in this
case follows directly from (\ref{zknf}) but the angular dependence,
which is now on $S_5/\zet_p$, requires trickier manipulations due to 
the non-commutativity of the $SO(6)_{\cal R}$ isometry group.    
The new ansatz can be written as
\bea
\chi_{a}&=&r g_{ab}({\bf e}_b+\hat{\chi}_b)\nonumber\\
\zeta_{\dalpha}^{A} &=& \bar{{\bf \eta}}_{\dalpha}^{A}
+\rho^{-{1\over 2}}
\hat{\zeta}_{\dalpha}^{A}\nonumber\\
{\cal M}_{\alpha}^{A} &=& {\bf \xi}_{\alpha}^{A}
+ \rho \hat{a}_{\alpha \dalpha}\bar{{\bf \eta}}_{\dalpha A}+ 
\rho^{{1\over 2}}\hat{\cal{M}}_{\alpha}^{A}
\label{spl} \ \ ,
\eea
where we denoted by $r g_{ab} {\bf e}_b$ the multi-instanton
center of mass degrees of freedom,
which are now no longer diagonal. Instead,
they are built out of square identity blocks:
\bea    
{\bf e}_{AB}&\equiv& 
{1\over \sqrt{8}} {\bf e}_b\Sigma^b_{AB}=
e_{AB}\delta_{i_{q+q_A} j_{q-q_B}} \ \ , \nonumber\\ 
\bar{{\bf \eta}}_{\dalpha}^{A}&\equiv&\bar{\eta}_{\dalpha}^{A} 
\delta_{i_q j_{q+q_A}} \ \ , \nonumber\\
{\bf \xi}_{\dalpha}^{A}&\equiv& 
{\xi}_{\dalpha}^{A} \delta_{i_q j_{q+q_A}}  \ \ ,
\label{block}
\eea  
suitably disposed inside the $K\times K$ matrix
in a way which preserves the $\Gamma$-invariance \cite{hk}. 
We have introduced in (\ref{spl}) a reference 
unit vector $e_b=-\bar{\Sigma}_b^{AB}e_{AB}/{\sqrt 2}$ 
which we fix 
%parametrize a point in $S^5$. By exploiting $SO(6)$ rotational invariance,
%we can choose this vector 
to lie along, say, the first axis,
$e_b=\delta_{b1}$.
Moreover, $g_{ab}$ is a matrix 
in the coset group $SO(6)/SO(5)$ parametrizing the orientation 
with respect to $e_b$ of the multi-instanton center of mass,
{\it i.e.} a position in $S^5$. 
Now, by plugging (\ref{spl}) 
in the action appearing in (\ref{zkn4}) and reabsorbing the remnant dependence 
on $g_{ab}$ of the Yukawa term by a suitable redefinition of the Clifford 
gamma matrices, one can easily check the factorization of a center of mass
term $AdS_5\times S^5/\zet_p$. 
The global $\zet_p$ identification 
is the result of a remnant discrete symmetry 
${\bf e}_{AB}\sim e^{2\pi i (q_A+q_B)/ p} {\bf e}_{AB}$ \cite{hkm}.

Notice that there is no deformation of the $AdS$ geometry.
Indeed, in this case $k_0=k_1=...=k_{p-1}$, which implies 
$\hat{k}_q=k_q$, and from (\ref{kb}) one can immediately see
that no additional powers of $r$ come
into the measure (\ref{zknf}).
 We conclude that regular instantons probe an
$AdS_5\times S^5/\zet_p$ (just as in the conformal case \cite{hkm})
even in the presence of fractional D3-branes, i.e. non conformal geometries.
This is again in agreement with our field theory expectations since
the overall sum of the gauge coupling constants, related to the
ten-dimensional type IIB coupling constant, does not run.

\section{D-instanton actions and topological field theories}

It is very useful for practical computations to write down a topological
version of the multi-instanton action we discussed in the previous
section. In fact the presence in this case of a scalar supersymmetry $Q$
allows to display interesting localization properties of the functional
integral. Moreover, the Ward identities associated
to $Q$ allow in some cases to easily rule out the contribution of
$Q$-exact terms in the action \cite{topo,dhk,nek1}.
In \cite{topo} this task was accomplished starting from the topologically twisted
version of ${\cal N}=2$ SYM$_4$ in $\real^4$. In this model, the functional integration
over the fields can be done exactly, and localizes 
the partition function on the 
instanton moduli space \cite{twist};
correspondingly, the dynamics of the multi-instanton collective coordinates 
in presence of a non-trivial vacuum expectation value for the
scalar field is described by a cohomological action on the ADHM variables.
 
The introduction of auxiliary fields
and the D-brane interpretation leading to (\ref{Sd}), in conjunction with
the orbifold projection, allows a more general formulation of the above
cited results. It is in fact now possible to twist the theory at the level
of the parameters describing the moduli space and to show that the theory
is given by the implementation of the ADHM constraints {\it \`a la} BRST.
All previous results are instantly recovered: the results with a vacuum
expectation value for the scalar in the theory are obtained by shifting 
the $\chi_a$ fields surviving the orbifold projection. Also the topological
theory obtained in the ${\cal N}=4$ case \cite{dhk} matches with our results 
after the orbifold projection.

We remark that the partition function we considered in the previous section 
is multiplied by the eight fermionic zero modes
$(\xi_{\alpha}^{\dA},\bar{\eta}_{\dalpha}^{\dA})$
associated to the translational and superconformal invariance,
such that, in order to get well defined physical quantities, 
one is forced to insert suitable observables.
The presence of a vacuum expectation value for the
scalar fields breaks superconformal invariance and makes the associated 
fermionic zero-modes
$\bar{\eta}_{\dalpha}^{\dA}$ to appear in the multi-instanton action.
Correspondingly, the partition function of the {\it centered} moduli space
\footnote{We recall that the centered moduli space 
is defined 
as the quotient of the usual moduli space 
with respect to the instanton center of mass $x_m$
and the associated four fermionic zero-modes $\xi_{\alpha}^{\dA}$.} 
becomes itself 
a non-trivial observable.
However, gauge theories with a non trivial v.e.v. correspond to less symmetric 
D-branes configurations, making the explicit computations
much more involved. In this case, the localization
properties which show up in the topological formulation
could give useful simplifications. On the other hand, this
formulation could be
used to identify a set of non-trivial observables useful
to analyse finite $N$ effects also in the conformal phase.
We finally remark that another way to break superconformal invariance 
is to consider a non-commutative geometry background; as we will see,
also this model can be easily included in a topological framework.

The ${\cal N}=2$ D-instanton action can be written in the 
cohomological theory set up by resorting to the standard twisting procedure
\cite{twist}.
In fact, as we discussed in Sect.3, after the orbifold 
projection (\ref{Ginv}), the field content of the theory
can be read from (\ref{aM}) by restricting $A,B$ to the
$\dA,\dB$ indices labelling the fundamental representation of the
${\cal N}=2$ automorphism group $SU(2)_{\dA}$.
%, since the original internal group
%of the ${\cal N}=4$ theory,
%$SU(4)$, has been projected to $SU(2)_A\times U(1)$.
We then redefine the four dimensional rotation group $SU(2)_L\times SU(2)_R$
acting on the projected fields
as $SU(2)_L\times SU(2)_R^\prime$, where $SU(2)_R^\prime$ is the diagonal
subgroup of $SU(2)_R\times SU(2)_{\dA}$.
We remark that on a flat manifold
this simply amounts to a redefinition of variables, such that the
conventional and the twisted formulation are completely equivalent.

In view of the above considerations, we also consider the possibility of 
turning on a vacuum expectation value $v$ for the adjoint scalar and a 
background B-field along the D3-brane directions. 
The former correspond to a separation between the D3-branes and is 
easily realized by replacing $\phi$ by $\phi+v$ in all previous formulas
\footnote{For simplicity we restrict $v$ to lie in the Cartan
subalgebra of the gauge group.}. The latter can be incorporated by adding a 
Fayet-Iliopoulos term $\zeta^c$ to the D(-1)-D3 lagrangian
and give rise
to a theory on a non--commutative space--time \cite{sw1}
\be
[x_m,x_n]=-i\zeta^c\bar{\eta}^c_{m n} \ \ .
\label{nc}
\ee
Accordingly, the bosonic ADHM constraints are deformed by the
parameter $\zeta^c$ \cite{nek2}, such that we have, in terms of the
twisted variables (where $\dot{A}$ is identified with $\dalpha$)
\bea
&&\bar{w}^{\dot{\alpha}} \mu_{\dot{\alpha}}
-\bar{\mu}^{\dot{\alpha}} w_{\dot{\alpha}}
-2[a_n,{\cal M}_n]=0 \ \ , \nonumber\\
&&\bar{w} \tau^c \mu +\bar{\mu} \tau^c w
    - 2i \bar{\eta}_{mn}^c[a_m,{\cal M}_n]=0 \ \ , \nonumber\\
&& \bar{w} \tau^c w-i \bar{\eta}_{mn}^c [a'_{m},a'_{n}]=\zeta^c \ \ .
\label{adhm-twist}
\eea
To these constraints we associate the doublets of auxiliary fields $(\lambda_c, D_c)$ and
$(\bar{\phi}, \eta)$, transforming under the scalar supercharge $Q$ as
\bea
Q\, \lambda_c&=& D_c \quad\quad  ~~Q\, D_c=[\phi, \lambda_c] \ \ , \nonumber\\
Q\, \bar{\phi}&=&\eta \quad\quad  ~~~~ Q\,\eta=[\phi,\bar{\phi}] \ \ ,
\nonumber\\
Q\, \phi&=&0  \quad\quad  ~~~~~~Q\, v= Q \bar{v}=0\nonumber\\
Q\, \zeta^c&=& 0 \ \ .
\label{aux}
\eea
where $\phi$ is the generator of $U(k)$ transformations under which
(\ref{adhm-twist}) are invariant.

The cohomological action which
implements the constraints (\ref{adhm-twist}) reads
\bea
S &=& {\rm tr}_k\, Q\Big[ (\bar{\phi}+\bar{v}) (\bar{w}^{\dot{\alpha}}
\mu_{\dot{\alpha}}
-\bar{\mu}^{\dot{\alpha}} w_{\dot{\alpha}}-2[a_n,{\cal M}_n]) + {1\over{g_0^2}}
\eta[\phi,\bar\phi] + \nonumber\\
&& + \lambda_c \left(\bar{w} \tau^c w - i \bar{\eta}^c_{mn} [a_{m},a_{n}] - \zeta^c \right)
   - {1\over{g_0^2}}\lambda_c D_c \Big] \ \ ,
\label{coho}
\eea
with
\bea
Q\, a_n &=&{\cal M}_n \quad\quad   Q \,{\cal M}_n=[\phi, a_m] \nonumber\\
Q\, w_{\dot{\alpha}}& =&\mu_{\dot{\alpha}} \quad\quad
~~Q\, \mu_{\dot{\alpha}}= - w_{\dot{\alpha}}(\phi+v) \nonumber\\
Q\, \bar{w}^{\dot{\alpha}}&=&\bar{\mu}^{\dot{\alpha}} \quad\quad
~~Q\, \bar{\mu}^{\dot{\alpha}}= (\phi+v) \bar{w}^{\dot{\alpha}} 
\label{susy}
\eea
We observe that in this context $g_0^2$ can be understood as a
"gauge--fixing" parameter, which can be continuously deformed without
affecting the results as long as the $Q$-invariant
physical observables we are computing
are well defined. In particular, we can send
$g_0^2\rightarrow\infty$, such that the ADHM constraints
(\ref{adhm-twist}) are implemented as Dirac delta-functions instead to
be spread out as Gaussian weights in the functional integral
\cite{nek1,dhk}. As we have seen in the previous sections, this
significantly simplifies the explicit computations.

In the presence of $N_F$ fundamental hypermultiplets,
the multi-instanton action gets another term \cite{hkm} which can be
written as
\be
S_{hyp} = - Q {\rm tr}_k \left[ \tilde{h}{\cal K} + \tilde{{\cal K}} h \right] \ \ ,
\label{hyp}
\ee
where $({\cal K},\tilde{{\cal K}})$ are
respectively $k\times N_F$ and $N_F\times k$ matrices denoting the
collective fermionic coordinates for the fundamental matter fields and
$(h,\tilde{h})$ are bosonic auxiliary variables, transforming as
\bea
Q\, {\cal K}&=& h \quad\quad Q\,h = \phi{\cal K} \ \ , \nonumber\\
Q\, \tilde{{\cal K}}&=& \tilde{h} \quad\quad Q\,\tilde{h} = -\tilde{{\cal K}} \phi \ \ .
\label{Qhyp}
\eea

By acting with $Q$ on the l.h.s of (\ref{coho}) and splitting
the action as in (\ref{cometipare}) we get, after integration on
the auxiliary variables $(h, \tilde{h})$
\bea
S_{G} &=& {\rm tr}_{k}\big( [\phi,\bar{\phi}]^2
-\eta [\phi,\eta]+\lambda_c[\phi,\lambda_c]
-D^{c}D^{c}\big) \, ,
\nonumber\\
%\label{Sg2} \\
S_{K} &=&{\rm tr}_{k}\big(-2[\phi,a_{n}][\bar{\phi},a_{n}]
+\left[(\phi+v)(\bar{\phi}+\bar{v}) + h.c. \right]
\bar{w}^{\dot{\alpha}} w_{\dot{\alpha}}\nonumber\\
&&+2 {\cal M}_m[\bar{\phi},{\cal M}_m]+2
 (\bar{\phi}+\bar{v})\bar{\mu}^{\dot{\alpha}}\mu_ {\dot{\alpha}}\big) ,
%\label{Sk2} \\
\nonumber\\
S_{D} &=& {\rm tr}_k\,\big[ \eta (\bar{w}^{\dot{\alpha}} \mu_{\dot{\alpha}}
-\bar{\mu}^{\dot{\alpha}} w_{\dot{\alpha}}
-2[a_n,{\cal M}_n] )
-\lambda_c\left(\bar{w} \tau^c \mu
+\bar{\mu} \tau^c w - 2i \bar{\eta}_{mn}^c[a_m,{\cal M}_n]\right)
\nonumber\\
&&+ D^{c}\left(\bar{w} \tau^c w - i \bar{\eta}^c_{mn} [a_{m},a_{n}]
               -\zeta^c \right)\big]  \nonumber\\
S_{hyp}&=& {\rm tr}_k \left[2 \tilde{{\cal K}}\phi{\cal K}\right] \ \ .
\label{Sd2}
\eea
We remark that this action can also be obtained from dimensional
reduction to $(0+0)$ dimensions of the two dimensional $(0,4)$ supersymmetric
Yang--Mills theory which describes the low-energy dynamics of
a $D1-D5$ system wrapping a $\real^4/\zet_2$ space. In this case, the 
fermionic symmetry $Q$ is given by a suitable combination of the 
supersymmetry charges \cite{w2}.

Few comments are in order for the pure ${\cal N}=2$ theory.
When $N_F=0$, the action (\ref{Sd2}) coincides for $v=\zeta^c=0$ with the
multi-instanton action of the previous section (2.1) written in terms of
the twisted variables
\bea
\lambda^{\dalpha}_{\dA} &=&
(\sqrt{2}\pi i)^{1\over 2}(i \eta\sigma^4+
\lambda_c \sigma^c)^{\dalpha}_{\dA    }\nonumber\\
{\cal M}^{\alpha \dA    } &=&
({\pi i \over \sqrt{2}})^{1\over 2} {\cal M}_n \bar{\sigma}_n^{\alpha\dA}
\nonumber\\
\mu^{\dA} &=&({\pi i \over \sqrt{2}})^{1\over 2}\mu^{\dA}
\eea
Moreover, we recall that a cohomological action for the commutative case
$\zeta^c=0$ and in presence of a
non-trivial v.e.v. $v$ for the scalar field was obtained in \cite{topo} by
resorting to the topologically twisted formulation of the ${\cal N}=2$ SYM
theory. Once the ADHM constraints have been enforced by integration on the
lagrangian multipliers $(D^c,\lambda^c,\eta)$,
the same multi-instanton action
can be obtained from the $S_K$ term in
(\ref{Sd2}) by integrating on $(\phi,\bar\phi)$,
which implements the equation of motion for the scalar field
\be
L\phi=[{\cal M}_m,{\cal M}_m] - \bar\mu^{\dalpha}\mu_{\dalpha}  \ \ ,
\label{phieom}
\ee
where the operator $L$ was defined in (\ref{L}).
The scalar supersymmetry operator (\ref{susy}) coincides then
with the covariant derivative on the moduli space defined in
\cite{topo}
\footnote{actually, in \cite{topo} the $Sp(1)\sim SU(2)$ ADHM formalism has
been adopted; hence, the ADHM auxiliary group is $O(k)$. This makes
to appear sligthly different notations for the explicit expression of
the action and the connection, but obviously the geometrical
interpretation holds unaltered.}
,
\be
Q = s + {\cal C}
\label{c}
\ee
where $s$ and ${\cal C}$ are respectively the exterior derivative and the
$U(k)$ connection on the ADHM moduli space, $\phi$ being the corresponding
$U(k)$ curvature.

In Appendix B we show how our ${\cal N}=2$ topological theory can be obtained
from that derived in \cite{dhk} for the ${\cal N}=4$ case by 
using the same projection procedure discussed in section 3.

\section{Instantons on ALE spaces}

Finally we would like to comment on the case 
in which the orbifold quotient is taken along the directions longitudinal
to the D3-brane system. These directions form a $\real^4$ space acted upon by
the Lorentz group $SO(4)\cong SU(2)_L \times SU(2)_R$.  
This case has been extensively studied in \cite{dmjm} and
D-instantons were shown to reproduce the Kronheimer-Nakajima
construction of self-dual connections for gauge theories 
living on ALE spaces \cite{kronnak}.
The projection is similar to the one performed
in section 2 but now $\Gamma$ is acting on the Lorentz indices $\alpha,\beta$
of $SU(2)_L$.

There is however an important difference respect to our 
previous considerations of gauge instantons in $\real^4$. 
Indeed, noticing that the ALE space geometry support topologically
non-trivial two cycles, the instanton solution is now classified 
by both the first and second Chern classes.
More precisely one can associate \cite{Bianchi:1996zj}
a tautological bundle ${\cal T}$ with fiber the regular
representation $R$ of $\Gamma$ and base the
ALE space itself. Under the action of $\Gamma$ this tautological
bundle admits a decomposition ${\cal T}=\sum_q {\cal T}_q \otimes R_q$
with $R_q$ ($q=0,1,....p-1$) the irreducible representation of
$\Gamma$. The first Chern Class $c_1({\cal T}_q)$ of the ${\cal T}_q$
bundles, $q\neq 0$
( $c_1({\cal T}_0)=0$ ), form a basis of the second cohomology group
and satisfy
\be
\int_{\rm ALE} c_1({\cal T}_q)\wedge c_1({\cal T}_{q'})
=(C^{-1})_{q q'}
\ee   
with $C^{-1}$ the inverse of the Cartan matrix of the unextended 
Dynkin diagram.
The restriction to the interesting case of 
instanton solutions with vanishing first Chern
class impose, as we will see, strong constraints on the allowed instanton
configurations $\{ k_q \}$ for a given partition $\{ N_q \}$. 
%Differently from the construction described in section 3, 
%the topological properties of the gauge connection and of the base
%manifold specify completely the values of the quantities $k_q, N_q$ introduced
%earlier. 
In the following we follow closely \cite{Bianchi:1996zj} 
whose notation we adapt to 
the computations carried out in the previous sections.

Once again the starting point is the $K=\sum_q k_q$
instanton solution of ${\cal N}=4$ SYM$_4$ on a
flat space and gauge group $U(N)$, with $N=\sum_q N_q$.
The projection (\ref{Ginv1}) is now replaced by
\bea
w_{\dalpha} &=&\gamma_N \, w_{\dalpha} \, \gamma_K^{-1}
\quad\quad 
\mu^A =  \gamma_N \, \mu^A
\,\gamma_K^{-1} \nonumber\\
\bar{w}^{\dalpha} &=&\gamma_K \, \bar{w}^{\dalpha} \, \gamma_N^{-1}
\quad\quad
\bar{\mu}^A =  \gamma_K \, \bar{\mu}^A
\,\gamma_N^{-1} \nonumber\\
a_{\alpha\dalpha} &=& e^{2\pi i {q_{\alpha} \over p}}
\gamma_K \, a_{\alpha\dalpha}\, \gamma_K^{-1}\quad\quad
{\cal M}^A_{\alpha} = e^{2\pi i {q_{\alpha} \over p}}
\gamma_K \, {\cal M}^A_{\alpha}\, \gamma_K^{-1}\label{Ginv1}\\
\chi_{AB} &=& 
\gamma_K \, \chi_{AB}\, \gamma_K^{-1}\quad
D^c = \gamma_K \, D^c\, \gamma_K^{-1}\quad
{\lambda}^A_{\dalpha}= 
\gamma_K \, {\lambda}^A_{\dalpha} \, \gamma_K^{-1}
\nonumber
\eea
with $q_1=-q_2=1$, and
$\gamma_K$, $\gamma_N$ the $K\times K$ and $N\times N$ matrices
realizing the orbifold
group action on the Chan-Paton indices. Notice that supersymmetry 
is preserved by this projection since $\Gamma$ acts in the same way
on the different components of a given supermultiplet.

The surviving components read 
\bea
w_{\dalpha}^q &=& w_{\dalpha i_q u_q}\quad\quad
\mu^A_{q} = \mu^A_{i_{q} u_q} \nonumber\\
a^{\alpha\dalpha}_q &=& a^{\alpha\dalpha}_{i_{q+q_\alpha} j_q}\quad\quad
{\cal M}^{A\alpha}_q = {\cal M}^{A\alpha}_{i_{q+q_\alpha} j_q} \nonumber\\
\chi^{AB}_q &=& \chi^{AB}_{i_{q} j_{q}} \quad
D^c_q= D^c_{i_q j_q}
\quad
\lambda^{A\dalpha}_q = \lambda^{A\dalpha}_{i_{q} j_q} \ \ .
%\nonumber\\
%D^c_q &=& D^c_{i_q j_q}
\label{invale}
\eea
The moduli space of multi-instanton solutions
and ADHM constraints can be described as before 
through (\ref{aM}), (\ref{constr}) but now in terms of the
invariant components (\ref{invale}).
In particular the dimension of the moduli space is 
given by the total number of components in the first two lines
of (\ref{invale}),
minus the number of ADHM constraints (given by  
${\rm dim}\, D^c=3 k_q^2$ for each $q$),
minus the dimension of the auxiliary gauge group $\prod_q U(k_q)$: 
\bea
{\rm dim}\, {\cal M}_B &=& 4\sum_q (k_q N_q+\hat{k}_q k_q-k_q^2) \nonumber\\
{\rm dim}\, {\cal M}_F &=& 8\sum_q (k_q N_q+\hat{k}_q k_q-k_q^2)
\label{dimension}
\eea
where $\hat{k}_q={1\over 2} \sum_{\alpha} k_{q+q_\alpha}$.
The resulting dimensions in (\ref{dimension}) are in agreement 
with \cite{Bianchi:1996zj}.
  
The definition of the ADHM matrices 
(\ref{del}) and the ADHM ansatz (\ref{An}) require instead
some generalization in order to encompass the non-triviality
of the ALE geometry. This can be easily done following
\cite{Bianchi:1996zj}. We first notice that the center of mass position 
$x_m$ of the multi-instanton solution can always be defined
as the trace part of the $a_m$ matrix, as in section 3, see (\ref{splitting}). 
From (\ref{invale}) one can then easily see that
$x_m$ parameterize a point in the $\real^4/\zet_p$ space    
if we test the geometry with regular D-instantons.
If instead we use fractional D-instanton probes,
$x_m$ is simply frozen to zero.

In order to extend the ansatz (\ref{An}) to the ALE situation
we should first give a covariant definition 
for $b$ in (\ref{bs})  
\be
b=\sigma^n_{[2]\times [2]}{\bf 1}_{[k]\times [k]}\nabla_n \Delta
\quad\quad
\bar{b}=\bar{\sigma}^n_{[2]\times [2]}{\bf 1}_{[k]\times [k]}
\nabla_n \bar{\Delta} \ \ ,
\label{bc}
\ee
where $\nabla_n=\partial_n + A_n^{\cal T}$ is the covariant
derivative respect to a connection $A_n^{\cal T}$ in 
the tautological bundle ${\cal T}$.
The ADHM ansatz for the gauge potential (\ref{An}) and adjoint 
fermionic zero modes are modified now to \cite{Bianchi:1996zj,Carpi:1999ig}
\bea
A_{n} &=&U^\dagger\nabla_{n} U \label{Ae}\\
\Psi^A &=& \bar{U}\left({\cal M}^A f \bar{\sigma^n}\nabla_n \bar{\Delta}
-\nabla_n \Delta \sigma^n f \bar{{\cal M}}^A\right)U 
\eea
which covariantly generalize (\ref{An}).
The matrices $U$ and $\bar{U}$ are defined as before 
as the kernels of $\Delta,\bar{\Delta}$.
Moreover, the ADHM matrices decompose under
$\Gamma$ as a collection of maps 
\be
\Delta^q_{[N_q+2k_q]\times [2\hat{k}_q]}:
W_q+(V_q\otimes {\bf 1}_{2\times 2})\rightarrow Q\times V_q  \ \ .
\ee
These properties can be used to relate the Chern character of the 
instanton bundle
to the Chern characters of the individual bundles
${\cal T}_q$.
The instanton bundle 
is specified then by giving the first and second Chern class 
\bea
c_1&=&\sum_q \left( N_q+2k_q -2\hat{k}_q \right)
c_1({\cal T}_q)\nonumber\\
c_2&=&\sum_q \left( N_q+2k_q -2\hat{k}_q \right)
c_2({\cal T}_q)+\frac{K}{|\Gamma|}\label{c2}
\eea
In particular an instanton solution with vanishing first Chern class is given
by  
\be
N_q+2k_q-2\hat{k}_q=0 \quad\quad {\rm for}~~~q>0.
\label{c10}
\ee 
This is a highly non trivial constraint on the allowed values of $(k_q,N_q)$.
Notice that only in this case the instanton number defined as
$K/|\Gamma|=K/p$ coincides with the second Chern class.  
In \cite{Bianchi:1996zj} the reader will  be able to find detailed
discussions of various cases. Here we simply recall the results in the
simplest context: the $SU(2)$ gauge bundle on the Eguchi-Hanson
blown down space $\real^4/\zet_2$. Solutions to (\ref{c10}) in this case
are given by either $\vec{N}=(2,0), \vec{k}=(k,k)$ or   
$\vec{N}=(0,2), \vec{k}=(k-1,k)$. They lead 
to instanton solutions with integer and half-integer second Chern
class respectively, as can be easily see from (\ref{c2}).  
The dimension of the multi-instanton moduli space can be read from 
(\ref{dimension}) and 
turns out to be respectively equal to $8k$ and $8k-4$,
in agreement with  \cite{Bianchi:1995ad}.
A computation of the partition function for the lowest value of the
Chern class, $c_2=1/2$, was carried out in \cite{Bianchi:1995ad}
yielding the bulk contribution to the Euler number of the 
moduli space\footnote{It is known and rigorously proven that for
$c_2=1/2$ the moduli space is a copy of the base manifold (the 
Eguchi-Hanson manifold) whose bulk contribution is $3/2$.}.   

Having described the main topological properties of the 
instanton solutions described by (\ref{invale}),
we come back to the study of the moduli space geometry. 
The low energy effective action (\ref{Sd}) 
written in terms of (\ref{invale}) gives again a
very tractable
description of the physics around the multi-instanton background.
Since the whole $SO(6)$ ${\cal R}$-symmetry is clearly
preserved by the projection, this action describes
instantons in an ${\cal N}=4$ gauge theory,
living on an ALE space.

It is a straightforward exercise to
apply the techniques of the previous section to show that
for a regular instanton probe
the center of mass degrees of freedom factor out from the
multi-instanton measure and describe a point in an
euclidean 
$S^5\times AdS^E_5/\zet_p$ space,  
with $\zet_p$ acting on the four-dimensional
$\real^4$ boundary of the $AdS^E_5$. 
In the case of fractional instantons
only the origin of this boundary space 
%in the boundary of $AdS_5\times S^5$ (not to be confused
%with the boundary of $AdS$ itself) 
is clearly probed. 
The crucial observation is that, unlike
in the non conformal situations considered in the previous section, no
$r$ dependence comes now from the determinants
$({\rm det}_{2k_q} W {\rm det}_{4k_q} {\chi}) \sim y^{4k_q}$
in (\ref{zkn4}), since the orbifold projection acts in the
same way on the $W$ and $\chi$ fields, balancing their contribution
against each other.
On the other hand also the various contributions
to the measure coming from the Jacobian of the scalings
(\ref{splitting}) cancel, leaving a conformal
$AdS^E_5/\zet_p\times S_5$ geometry.

\vskip 0.5in

{\bf Acknowledgements}

We acknowledge helpful discussions with L. Baulieu, G. Bonelli, V.V. Khoze,
C. Johnson, M. Petrini, M. Porrati, R. Russo, M. Trigiante.
This work was supported in part by the EEC contracts 
HPRN-CT-2000-00122, HPRN-CT-2000-00131, by 
the INTAS contract 55-1-590 and by the COFIN project of MURST.

\appendix
\section{Appendix }
In section 2 we have seen how, by modding the internal symmetry group,
the number of supersymmetric charges is reduced and we have discussed
the case of a  ${\cal N}=4$ theory going to a ${\cal N}=2$.
In section 4 this latter theory has been twisted allowing a topological 
interpretation. The following material   
will help the reader to convince
himself that in this reduction the right quantum numbers for the ${\cal N}=2$
theory are generated.

To avoid the 
introduction of more formalism, we will use the notations of \cite{georgi}, to
which we refer the reader for a complete treatment of the subject.
As it is well known, the highest weight of a representation, 
$\mu=\sum q^i\mu^i$, can be built as a tensor product starting from the 
fundamental weights $\mu^i$. The representations whose weights are the $\mu^i$
are the fundamental representations $D^i$. For $SO(10)$ we concentrate
on two such
representations, $D^4$ and $D^5$ with weights $\mu^4=(e^1+\ldots -e^5)/2$ and
$\mu^5=(e^1+\ldots +e^5)/2$, where the $\pm e^i\pm e^j, i\ne j$ 
are the roots of the algebra.
Both representations are 16 dimensional and if we denote by 
$1/2\sum_i\eta_i e^i$
 a generic weight (with $\eta^i=\pm 1$), 
we have that for $D^4$, $\prod_i\eta_i=-1$,
while for $D^5$, $\prod_i\eta_i=1$. This means that $D^4$ ($D^5$) has 
negative (positive) parity, since the $\eta_i$ can be chosen to be the 
eigenvalues of some $\sigma^i_3$ Pauli matrices out of which the Dirac 
$\Gamma_{11}$ matrix (in the Clifford representation)
can be built as the tensor product $\Gamma_{11}=\otimes_{i=1}^5 \sigma_3^i$.
Let us now see what happens to $D^4, D^5$ under the regular maximal subgroup,
{\it i.e.} the subgroup whose rank is the same as that of the original group.
These subgroups are obtained by studying the extended $\Pi$ system of the 
algebra. In our case we obtain 
$SO(10)\to SO(6)\times SO(4)\cong SU(4)\times SU(2)_L\times SU(2)_R$.
The simple roots of $SO(6)$ are now
\bea
\alpha^1&=&e^1-e^2\nonumber\\
\alpha^2&=&e^2-e^3\\
\alpha^0&=&-e^1-e^2\nonumber
\label{srootsso6}
\eea
and those of the $SU(2)_L\times SU(2)_R$ are
\bea
\alpha^4&=&e^4-e^5\nonumber\\
\alpha^5&=&e^4+e^5
\label{srootsso4}
\eea
The weights of $D^5$ are now divided into two sets 
$\{\eta_1\eta_2\eta_3=1, \eta_4\eta_5=1\},
\{\eta_1\eta_2\eta_3=-1, \eta_4\eta_5=-1\}$.
Recapitulating, we have seen that $D^5\to ({\bf 4},{\bf 2},{\bf 1})\oplus 
(\bar {\bf 4},{\bf 1},{\bf 2})$. Furthermore,
given that the maximal regular subgroup of $SU(4)\sim SO(6)$ is 
%$SU(2)\times SU(2)^\prime\times U(1)$, the fifteen generators of the
%rank three  $SU(4)$ algebra can be given in terms of Pauli matrices as
%$\{(\sigma_a)_{\alpha\beta}\otimes \delta_{ij},
%\delta_{\alpha\beta}\otimes (\sigma_a)^\prime_{ij},  
%(\sigma_a)_{\alpha\beta}\otimes  (\sigma_b)^\prime_{ij}\}$
%so that for the three Cartan generators of $SU(4)$, $H_i$,
%it is true that $H_1\subset SU(2), H_2 \subset SU(2)^\prime, 
%H_3\subset U(1)$.  
$SO(4)\times SO(2)\sim SU(2)\times SU(2)^\prime\times U(1)$
it follows that the $U(1)$ coming from the 
decomposition of the ${\bf 4}$ with positive parity $\eta_1\eta_2=+1$
has $U(1)$ quantum number $\eta_3=+1$. 

The positive parity Weyl-Maiorana spinor in ten space-time 
dimensions is
\be
\psi=\sqrt{\frac{\pi}{2}}\pmatrix{0\cr 1}\otimes
\pmatrix{{\cal M}^{ A}_\beta \cr 0}+
\sqrt{\frac{\pi}{2}}\pmatrix{1\cr 0}\otimes
\pmatrix{0 \cr\lambda^{ A}_{\dbeta} }
\label{spinor10d}
\ee
${\cal M}^{ A}_\beta, \lambda^{ A}_{\dbeta}$ are the $({\bf 4},{\bf 2},{\bf 1}
)$ and  $(\bar {\bf 4},{\bf 1},{\bf 2})$ parts respectively.

Let us now discuss the modding: we decide to label by $\mu=7,\ldots,10$ the 
directions longitudinal to our D3-brane.
The $\Gamma$ subgroup is
going to act on the transverse $\hat a=1, 2, 4, 5$ directions. 
We also consider $\Gamma=\zet_2$ to recover pure ${\cal N}=2$ SYM$_4$.
In this case the action of the $\zet_2$ is that of an inversion, $A$, which 
transforms
the spinor (\ref{spinor10d}) as $\psi^\prime(x^\prime)=\psi^\prime(A\cdot x)=
S(A)\psi(x)$. Requiring the invariance of the Dirac's equation we find
$S^{-1}\Gamma^{\hat a} S=-\Gamma^{\hat a}$ with $\hat a=1, 2, 4, 5$.
As it is well known, the solution to this equation is given by the parity 
matrix. 

A convenient representation of the ten dimensional Clifford algebra
is given by 
\be
\Gamma_a=\pmatrix{0 & \Sigma^a_{AB}\cr \bar\Sigma^a_{AB}& 0 \cr}\otimes 
\gamma^5 \qquad \Gamma_{\mu}= 1_{8\times 8}\otimes \gamma^\mu
\label{cliffalg}
\ee
where the $\gamma^\mu$'s define a four dimensional Clifford algebra
in the longitudinal directions
\be
\gamma^\mu=\pmatrix{0 & \sigma^\mu\cr \bar \sigma^\mu &0}\qquad
\gamma^5=\pmatrix{1 & 0\cr 0 & -1}
\ee
By using (\ref{cliffalg}) it is now easy to see that the condition of parity
invariance in the $\hat a=1, 2, 4, 5$ directions boils down to setting to zero
those fermionic components with $A=\bar{A}=3,4$. 
For what the vector is concerned, $\chi^\prime=A\cdot\chi=-\chi$ for
the components $\mu=1, 2, 4, 5$. 

Implementing these considerations in (\ref{Sd}) one easily obtains the measure
for the ADHM construction with ${\cal N}=2$. Modding out by a discrete subgroup
turns out to be a very effective way to deduce lagrangians with a lower number
of supersymmetries. 

\section{Appendix}
In section 3 we showed how to obtain a ${\cal N}=2$ supersymmetric theory
out of a ${\cal N}=4$ one by imposing $\Gamma$ invariance. The same exercise
can be carried on with topological theories. Our starting point will be the
${\cal N}=4$ topological theory studied in \cite{dhk}, which we rewrite here as
\be
S=  Q\,{\rm tr}_k\Big({1\over 4}\eta[\phi,\bar\phi]+
\vec H\cdot\vec\chi-
i\vec{\cal
E}\cdot\vec\chi-{1\over 2}\sum_{l=1}^6(\Psi^\dagger_l\bar\phi\cdot 
B_l+\Psi_l\bar\phi\cdot B_l^\dagger)\Big)\ ,
\label{qact}
\ee
where $\vec H = (H_{\dalpha}^{(a)}, H_{\dalpha}^{(f)}, i D^c)$ are the
auxiliary fields which implement the constraints 
$\vec{\cal E}= ({\cal E}_{\dalpha}^{(a)}, {\cal E}_{\dalpha}^{(f)},
{\cal E}^c)$ and $\vec{\chi}=(\chi_{\dalpha}^{(a)}, \chi_{\dalpha}^{(f)}, 
i \lambda^c)$ their fermionic superpartners, while
$B_l=(\phi^s,w_{\dalpha}, a_m)$ and $\Psi_l$ are the corresponding
fermionic superpartners.
The quiver projection of section 3.1 implies that the
$\phi^s$ fields
%, corresponding to the $y,\tilde y$ fields of \cite{dhk} 
are set to zero; then, 
%the constraints
%$({\cal E}_{\dalpha}^{(a)}, {\cal E}_{\dalpha}^{(f)})$ (see (4.5) of \cite{dhk})
%are trivially satisfied, and the corresponding auxiliary fields 
%$(H_{\dalpha}^{(a)}, H_{\dalpha}^{(f)})$ decouple from the action
%(\ref{qact}). Hence, dropping 
%$(\phi^s,H_{\dalpha}^{(a)}, H_{\dalpha}^{(f)})$ and their fermionic
%superpartners in (\ref{qact}), we are left with
also their
fermionic superpartners and the associated auxiliary fields 
are dropped, and we are left with  
\be
S=Q\, {\rm tr}_k \Big({1\over 4}\eta [\bar\phi,\phi]-\lambda_c D^c +
\lambda_c{\cal E}^c-{\cal M}_m[\bar\phi,a_m]+
{1\over 2}\bar\mu^{\dot\alpha}\bar\phi w_{\dot\alpha}+
{1\over 2}\mu^{\dot\alpha}\bar\phi \bar w_{\dot\alpha}\Big) \ \ ,
\label{formapp}
\ee
where ${\cal E}^c$ can be identified with the twisted bosonic constraint,
last equation in (\ref{adhm-twist}). 
Finally, by adding a vev to the scalar fields and making the simple rescalings 
$\phi\rightarrow 2\phi$, 
$\eta\rightarrow \eta/{g_0^2}$,
$D^c\rightarrow {D^c}/{g_0^2}$,
we get the action (\ref{coho}).


\begin{thebibliography}{999}


\bibitem{Maldacena:1998re}
J.~Maldacena,
%``The large N limit of superconformal field theories and supergravity,''
Adv.\ Theor.\ Math.\ Phys.\ {\bf 2} (1998) 231, hep-th/9711200.

\bibitem{fgpw} D.Z. Freedman, S.S. Gubser, K. Pilch and N.P. Warner,
%Renormalization Group Flows from Holography--Supersymmetry and a c-Theorem
Adv. Theor. Math. Phys. {\bf 3} (1999) 363, hep-th/9904017. 

\bibitem{kn} I. R. Klebanov and N. Nekrasov, Nucl.\ Phys.\ B574 (2000)
263, hep-th/9911096.

\bibitem{jpp}  C. V. Johnson, A. W. Peet and J. Polchinski
%Gauge Theory and the Excision of Repulson Singularities
Phys. Rev. {\bf D61} (2000) 086001, hep-th/9911161;\\
C.V. Johnson,
%Enhancons, Fuzzy Spheres and Multi-Monopoles
Phys. Rev. {\bf D63} (2001) 065004, hep-th/0004068;\\
A. Buchel, A. W. Peet and J. Polchinski,
%Gauge Dual and Noncommutative Extension of an N=2 Supergravity Solution
Phys. Rev. {\bf D63} (2001) 044009, hep-th/0008076 \\
N. Evans, C. V. Johnson and  M. Petrini,
%The Enhancon and N=2 Gauge Theory/Gravity RG Flows
JHEP {\bf 0010} (2000) 022, hep-th/0008081;\\ 
J. Polchinski,
%N = 2 Gauge/Gravity Duals
Int. J. Mod. Phys. {\bf A16} (2001) 707, hep-th/0011193;\\
M. Frau, A. Liccardo and R. Musto,
%The Geometry of Fractional Branes
Nucl. Phys. {\bf B602} (2001) 39, hep-th/0012035;\\ 
O. Aharony,
%A note on the holographic interpretation of string theory backgrounds 
%with varying flux
JHEP {\bf 0103} (2001) 012, hep-th/0101013;\\
M. Billo', L. Gallot and A. Liccardo,
{\it ``Classical geometry and gauge duals 
for fractional branes on ALE orbifolds''}, hep-th/0105258;\\ 
M.~Grana and J.~Polchinski,
{\it ``Gauge/Gravity Duals with Holomorphic Dilaton''},
hep-th/0106014.

\bibitem{bdflmp}
M.~Bertolini, P.~Di Vecchia, M.~Frau, A.~Lerda, R.~Marotta and I.~Pesando,
%``Fractional D-branes and their gauge duals,''
JHEP{\bf 0102} (2001) 014, hep-th/0011077.


\bibitem{prz}
M. Petrini, R. Russo, A. Zaffaroni,
{\it ''N=2 Gauge Theories and Systems with Fractional Branes''},
hep-th/0104026;\\
C.V. Johnson, R.C. Myers, A.W. Peet and S.F. Ross,
{\it ``The Enhancon and the Consistency of Excision''},
hep-th/0105077;\\
C.V. Johnson and R.C. Myers,
{\it  ``The Enhancon, Black Holes, and the Second Law''}
hep-th/0105159.

\bibitem{dhkmv} N. Dorey, T.J. Hollowood, V.V.
Khoze, M.P. Mattis and S. Vandoren,
% ``Multi-Instanton Calculus and the AdS/CFT Correspondence in N=4 Superconformal Field Theory''
Nucl.\ Phys.\ {\bf B552} (1999) 88, hep-th/9901128

\bibitem{bgkr} M. Bianchi, M.B. Green, S. Kovacs and 
G. Rossi, 
%{\it ``Instantons in supersymmetric Yang-Mills and D-instantons in 
%IIB superstring theory''},
JHEP {\bf 9808} (1998) 013, hep-th/9807033.

\bibitem{gns}
E.~Gava, K.~S.~Narain and M.~H.~Sarmadi,
%``Instantons in N = 2 Sp(N) superconformal gauge theories and the AdS/CFT  correspondence,''
Nucl.\ Phys.\ B {\bf 569} (2000) 183, hep-th/9908125.

\bibitem{hkm} T.J. Hollowood, V.V. Khoze and M.P. Mattis,
% ``Summing the Instanton Series in N=2 Superconformal Large N QCD''
HEP {\bf 9910} (1999) 019, hep-th/9905209.

\bibitem{hk} T. J. Hollowood and V.V. Khoze,
% ``ADHM and D-instantons in orbifold AdS/CFT duality''
Nucl.Phys. {\bf B575} (2000) 78, hep-th/9908035.

\bibitem{fsafm}
A. Fayyazuddin and M. Spalinsky, Nucl.\ Phys.\ {\bf B535} (1998)
219,hep-th/9805096 ; O. Aharony, A. Fayyazuddin and J. Maldacena,
{\bf JHEP} 9807 (1998) 013, hep-th/9806159.

\bibitem{kasi}
S. Kachru and E. Silverstein, Phys.\ Rev.\ Lett.\ {\bf 80} (1998)
4955, hep-th/9802183;\\
A. Lawrence, N. Nekrasov and C. Vafa, Nucl.\ Phys.\ {\bf B533}
(1998) 199, hep-th/9803015; M. Bershadsky, Z. Kakushadze and C.
Vafa, Nucl.\ Phys.\ {\bf B523} (1998) 59, hep-th/9803076;

\bibitem{aa} C. Angelantonj and A. Armoni, 
%Non-Tachyonic Type 0B Orientifolds, 
Phys. \ Lett.\ {\bf B482} (2000) 329; 
%Non-Supersymmetric Gauge Theories and Cosmological RG Flow
 Nucl. \ Phys.\ {\bf B578} (2000) 239, hep-th/9912257.

\bibitem{bimo} M. Bianchi and Jose F. Morales, 
%RG-flows and Open/Closed String Duality
JHEP {\bf 0008} (2000) 035,
hep-th/0006176;\\
{\it ``Anomalies, RG-flows and Open/Closed String Duality''},
Proceedings of the Ninth Marcel Grossman Meeting, Rome July 2-8, 2000,
hep-th/0101104.

\bibitem{adhm}
M.~Atiyah, V.~Drinfeld, N.~Hitchin and Yu.~Manin,
 Phys.\ Lett.\ {\bf 65A} (1978) 185.

\bibitem{sw}
{N.~Seiberg and E.~Witten, Nucl.\ Phys.\ {\bf B426} (1994) 19;
{\it ibid.} {\bf B431} (1994) 484.}

\bibitem{Dorey:1998xb}
N.~Dorey, V.~V.~Khoze and M.~P.~Mattis,
%``Supersymmetry and the multi-instanton measure,''
Nucl.\ Phys.\ B {\bf 513} (1998) 681, hep-th/9708036;
N.~Dorey, T.~J.~Hollowood, V.~V.~Khoze and M.~P.~Mattis,
%``Supersymmetry and the multi-instanton measure. II: From N = 4 to N = 0,''
Nucl.\ Phys.\ B {\bf 519} (1998) 470, hep-th/9709072.

\bibitem{Bruzzo:2000di}
U.~Bruzzo, F.~Fucito, A.~Tanzini and G.~Travaglini,
{\it ``On the multi-instanton measure for super Yang-Mills theories'',}
hep-th/0008225.

\bibitem{topo} D. Bellisai, F. Fucito, A. Tanzini and G. Travaglini,
% ``Instanton Calculus, Topological Field Theories and N=2 Super Yang-Mills Theories''
JHEP {\bf 0007} (2000) 017, hep-th/0003272; Phys.\ Lett.\ B {\bf 480} (2000)
365, hep-th/0002110.

\bibitem{dhk} N.Dorey, T.J. Hollowood and V.V. Khoze,
% The D-Instanton Partition Function''
JHEP{\bf 0103} (2001) 040,hep-th/0011247.

\bibitem{Fucito:1997ua}
N.~Dorey, V.~V.~Khoze and M.~P.~Mattis,
%``Multi-Instanton Calculus in N=2 Supersymmetric Gauge Theory,''
Phys.\ Rev.\ D {\bf 54} (1996) 2921, hep-th/9603136; ibid.\
%``Multi-instanton calculus in N = 2 supersymmetric gauge theory.  II: Coupling to matter,''
Phys.\ Rev.\ D {\bf 54} (1996) 7832, hep-th/9607202;
F.~Fucito and G.~Travaglini,
%``Instanton calculus and nonperturbative relations in N = 2  supersymmetric gauge theories,''
Phys.\ Rev.\ D {\bf 55} (1997) 1099, hep-th/9605215.

\bibitem{wttn} E. Witten,
Nucl.\ Phys.\ {\bf B460} (1996) 541, hep-th/9511030.

\bibitem{Douglas:1998uz}
M.~R.~Douglas,
{\it ``Branes within branes''}, hep-th/9512077;
%``Gauge Fields and D-branes,''
J.\ Geom.\ Phys.\ {\bf 28}, 255 (1998), hep-th/9604198.

\bibitem{Craps:1997gp}
B.~Craps, F.~Roose, W.~Troost and A.~Van Proeyen,
%``What is special Kaehler geometry?,''
Nucl.\ Phys.\ B {\bf 503} (1997) 565, hep-th/9703082;
L.~Andrianopoli, M.~Bertolini, A.~Ceresole, R.~D'Auria, S.~Ferrara, P.~Fre
and T.~Magri,
%``N = 2 supergravity and N = 2 super Yang-Mills theory on general scalar  manifolds: Symplectic covariance, gaugings and the momentum map,''
J.\ Geom.\ Phys.\ {\bf 23} (1997) 111, hep-th/9605032.

\bibitem{nek1}   G. Moore, N. Nekrasov and  S. Shatashvili,
% ``D-particle bound states and generalized instantons''
Commun.\ Math.\ Phys.\ {\bf 209} (2000) 77, hep-th/9803265.

\bibitem{dmjm}
M.~R.~Douglas and G.~Moore,
{\it ``D-branes, Quivers, and ALE Instantons''},
hep-th/9603167; C.~V.~Johnson and R.~C.~Myers,
%``Aspects of type IIB theory on ALE spaces,''
Phys.\ Rev.\ D {\bf 55} (1997) 6382, hep-th/9610140.

\bibitem{Maggiore:2001}
N.~Maggiore and A.~Tanzini, 
{\it ``Protected Operators in {\cal N}=2,4 Supersymmetric Theories''},
hep-th/0105005.


\bibitem{nsvzmrs}
V.A. Novikov, M.A. Shifman, A.I. Vainshtein and V.I. Zakharov,
Nucl.\ Phys.\ {\bf B229} (1983) 381;
T.R. Morris, D.A. Ross and C.T. Sachrajda,
Nucl.\ Phys.\ {\bf B264} (1986) 111.

\bibitem{twist} E. Witten,
% ``Topological Quantum Field Theory''
Commun.\ Math.\ Phys.\ {\bf 117} (1988) 353.

\bibitem{sw1} N.Seiberg and  E. Witten,
% ``String Theory and Noncommutative Geometry''
JHEP {\bf 9909} (1999) 032, hep-th/9908142.

\bibitem{nek2} N. Nekrasov and A. Schwarz,
% ``Instantons on noncommutative $\real^4$, and (2,0) superconformal six dimensional theory''
Commun.\ Math.\ Phys.\ {\bf 198} (1998) 689, hep-th/9802068.

\bibitem{w2} E. Witten,
% ``Phases of N=2 Theories in Two Dimensions''
Nucl.\ Phys.\ {\bf B403} (1993) 159, hep-th/9301042.

\bibitem{kronnak}
P.~Kronheimer and H.~Nakajima, Math.\ Ann.\ {\bf 288} (1990) 263.

\bibitem{Bianchi:1996zj}
M.~Bianchi, F.~Fucito, G.~Rossi and M.~Martellini,
%``Explicit Construction of Yang-Mills Instantons on ALE Spaces,''
Nucl.\ Phys.\ B {\bf 473} (1996) 367, hep-th/9601162.

\bibitem{Carpi:1999ig}
C.~Carpi and F.~Fucito,
%``Gauge zero-modes on ALE manifolds,''
Class.\ Quant.\ Grav.\ {\bf 16} (1999) 1805, hep-th/9809107.

\bibitem{Bianchi:1995ad}
M.~Bianchi, F.~Fucito, M.~Martellini and G.~Rossi,
%``Instanton effects in supersymmetric Yang-Mills theories on ALE gravitational backgrounds,''
Phys.\ Lett.\ B {\bf 359} (1995) 56, hep-th/9507003.

\bibitem{georgi} H. Georgi,
{\it ''Lie Algebras in Particle Physics''}, Benjamin, 1982.



\end{thebibliography}
\end{document}